\documentclass[a4paper,fleqn,usenatbib]{mnras}

\usepackage{newtxtext,newtxmath}

\usepackage[T1]{fontenc}
\usepackage{ae,aecompl}

\usepackage{multicol}
\usepackage{multirow}
\usepackage{mathtools}
\usepackage{graphicx}	
\usepackage{amsmath}	
\usepackage{lscape}     
\usepackage{longtable}  

\newcommand{\be}{\begin{equation}}
\newcommand{\ee}{\end{equation}}
\newcommand{\Msun}{M_\odot}
\newcommand{\maria}{}
\newcommand{\elena}{}
\newcommand{\secrev}{}

\newcommand{\thirdrev}{}

\title[Observations and modelling of SN~2017gpn]{Optical and spectral observations and hydrodynamic modelling of Type IIb Supernova~2017gpn}

\author[E.~A.~Balakina et al.]{
Elena~A.~Balakina,$^{1,2}$
Maria~V.~Pruzhinskaya,$^{2,3}$\thanks{E-mail: pruzhinskaya@gmail.com}
Alexander~S.~Moskvitin,$^{4}$
\newauthor Sergei~I.~Blinnikov,$^{2,3,5,6}$
Xiaofeng~Wang,$^{7,8}$
Danfeng~Xiang,$^7$
Han~Lin,$^7$
\newauthor Liming~Rui$^7$,
 and Huijuan~Wang$^9$ \\
$^{1}$ Lomonosov Moscow State University, Faculty of Physics, Leninskie Gory, 1-2, Moscow, 119991, Russia\\
$^{2}$ Lomonosov Moscow State University, Sternberg Astronomical Institute, Universitetsky pr.~13, Moscow, 119234, Russia\\
$^{3}$ Space Research Institute, 84/32 Profsoyuznaya Street, Moscow, 117997, Russia\\
$^{4}$ Special Astrophysical Observatory RAS, Nizhnij Arhyz, 369167, Russia\\
$^{5}$ NRC Kurchatov institute - ITEP, B.Cheremushkinskaya 25, 117218 Moscow, Russia\\
$^{6}$ Kavli IPMU, University of Tokyo, Kashiwa, 277-8583, Japan\\
$^{7}$ Physics Department and Tsinghua Center for Astrophysics, Tsinghua Unviersity, Beijing, 100084, China\\
$^{8}$ Beijing Planetarium, Beijing Academy of Science and Technology, Beijing, 100044, China\\
$^{9}$ National Astronomical Observatories, Chinese Academy of Sciences, Beijing, 100101, China\\
}

\date{Accepted XXX. Received YYY; in original form ZZZ}

\pubyear{2020}

\begin{document}
\graphicspath{{images/}}
\label{firstpage}
\pagerange{\pageref{firstpage}--\pageref{lastpage}}
\maketitle

\begin{abstract} 
In this work we present the photometric and spectroscopic observations of Type IIb Supernova~2017gpn. This supernova was discovered in the error-box of the LIGO/Virgo G299232 gravitational-wave event. We obtained the light curves in the $B$ and $R$ passbands and modelled them numerically using the one-dimensional radiation hydrocode \textsc{STELLA}. The best-fitting model has the following parameters: \maria{the} pre-SN star mass and \maria{the} radius are \thirdrev{$\rm M$~$\approx 3.5~M_\odot$ and $\rm R$~$\approx 50~R_\odot$}, respectively; \maria{the} explosion energy is \thirdrev{$\rm E_{exp} \approx 1.2\times10^{51}$~erg}; \maria{the} mass of radioactive nickel is \thirdrev{$\rm M_{^{56}Ni} \approx 0.11$}~$M_\odot$, which is completely mixed throughout the ejecta; and the mass of the hydrogen envelope \thirdrev{$\rm M_{H\_{env}}$~$\approx 0.06~M_\odot$}. Moreover, SN~2017gpn is a confirmed SN~IIb that is located at the farthest distance from the centre of its host galaxy NGC~1343 (i.e. \secrev{the projected distance is} $\sim$21~kpc). This challenges the scenario \maria{of} the origin of Type IIb Supernovae from massive stars.
\end{abstract}




\section{Introduction}
Type IIb Supernovae (SNe~IIb) are characterized by spectra evolving from dominant hydrogen lines at early times to increasingly strong helium features and progressively weaker hydrogen lines later on~\citep{1993Filippenko}. This is the reason why SNe~IIb are regarded as \maria{an} intermediate group between hydrogen-rich SNe~II and hydrogen-poor SNe~Ib. 
SNe~IIb \maria{are in the class of} the stripped-envelope core-collapse supernovae (CCSNe). It is supposed that progenitors of such supernovae are massive stars that have lost most of their hydrogen envelope \citep{1997Clocchiatti}. 

Nowadays there are two hypotheses explaining how stars can \elena{lose} the hydrogen envelope. The first scenario supposes the evolution of a rather massive $\rm M \simeq 25$~$\Msun$ single star with an average mass-loss rate of about $10^{-5}$~$\Msun$~per~year. Such a powerful stellar wind could provide the required outflow of hydrogen~\citep{1993Hoflich}. The second and more plausible scenario involves a mass transfer in a binary system where the progenitor star is a supergiant of moderate mass~\citep{1993Nomoto,1994Woosley,2012ApJ...757...31B}. The massive companion expands and fills its Roche lobe, after which mass transfer starts due to Roche-lobe overflow~\citep{2017Yoon}. 

Nevertheless, the progenitor nature of SNe~IIb is still not clear. While SNe~II form a continuous group as \cite{2014Anderson} and \cite{2015Sanders} established, \cite{2019Pessi} showed that SN~II light curves are distinct from those of SNe~IIb with no suggestion of a continuum distribution. This fact suggests that progenitors of SNe~IIb make up a separate group that is different from the SNe~II ones. However, it could also be a consequence of the lack of observational data: SNe~IIb make up less than 5~per~cent of all CCSNe according to the Open Supernova Catalog\footnote{\url{https://sne.space}}~\citep{SneSpace} and only about two dozen of them have detailed multicolour photometry appropriate for further study (including hydrodynamic modelling). 

To extend the sample of well-studied SNe~IIb, in this paper we present \maria{the} photometric and spectroscopic observations of SN~2017gpn. The photometry was performed with the Zeiss-1000 telescope~\citep{Zeiss} at the Special Astrophysical Observatory of the Russian Academy of Science (SAO~RAS). Spectroscopic data were obtained with \maria{the} Xinglong 2.16-m telescope at \maria{the} National Astronomical Observatory of China. Collected photometric data are used for the numerical light-curve (LC) calculations done by the radiation hydrocode \textsc{STELLA}~\citep{Blinnikov1998,Blinnikov2006}. These simulations \maria{give us} the parameters of the pre-supernova star and explosion characteristics. 

The interest in this supernova is also augmented by the fact that we usually only observe such supernovae in spiral galaxies in hydrogen-rich environments where young massive stars are being born~\citep{Filippenko1997}. In contrast to this, SN~2017gpn is located quite far from the active star-formation regions and \maria{the} spiral arms of the host galaxy. We also do not see any dwarf satellite galaxies at the SN \maria{location}. The unusual location of SN~2017gpn in \maria{the} host galaxy \maria{indicates} that the existing models of SN~IIb progenitors \maria{may} not explain all observational data and have to be reviewed.

This paper is organized as follows. In Section~\ref{sec:observations} we describe the observations, data processing, and resulting light curves and spectra. In Section~\ref{sec:modelling} we present the hydrodynamic modelling of SN~2017gpn and the parameters of the best-fitting model. Section~\ref{sec:discussion} contains a comparison of the modelling results, LC behaviour, and spectral features of SN~2017gpn with those for other SNe~IIb and a discussion of the unexpected location of SN~2017gpn relative to its host galaxy. Finally, we conclude the paper in Section~\ref{sec:conclusion}.

\section{Observations}
\label{sec:observations}
\subsection{Discovery}
On the last day of the second advanced detector observing run ``O2'', the LIGO/Virgo Collaboration released the G299232 alert\footnote{\url{https://gcn.gsfc.nasa.gov/other/G299232.gcn3}}. During the follow-up inspection of the gravitational-wave (GW) candidate error-box, on 2017 August 27.017 the MASTER Global Robotic Net~\citep{2010MASTER} discovered an optical transient named MASTER OT J033744.97+723159.0~\citep{GCN21719}.

On the discovery day, three spectra of MASTER OT~J033744.97+723159.0 were obtained with the ACAM instrument mounted on the William Herschel Telescope at La~Palma (Spain) by~\cite{GCN21737} and the analysis showed that the transient \elena{classifies as} SNe~IIb. Further observations on 2017 August~29 obtained with the SPRAT spectrograph on the Liverpool Telescope~\citep{GCN21755} and with the Xinglong 2.16-m telescope of \maria{the} National Astronomical Observatory of China~\citep{2017ATel10681....1R,2017ATel10684....1W} confirmed this classification by cross-correlating with a library of spectra with use of the Supernova Identification code (\textsc{SNID};~\citealt{2007ApJ...666.1024B}).
According to \textsc{SNID}, the spectrum with the highest correlation coefficient belongs to Type~IIb SN~1996cb at phase $-$2~d.

On 2017 September~6 at 03:21:12~UT,~\cite{2017Caimmi} reported the discovery of a supernova with the 0.24-m telescope from \maria{the} Valdicerro Observatory. The supernova received the IAU designation AT~2017gpn and was identified as MASTER OT~J033744.97+723159.0.

\begin{figure}
    \centering
	\includegraphics[width=0.75\columnwidth]{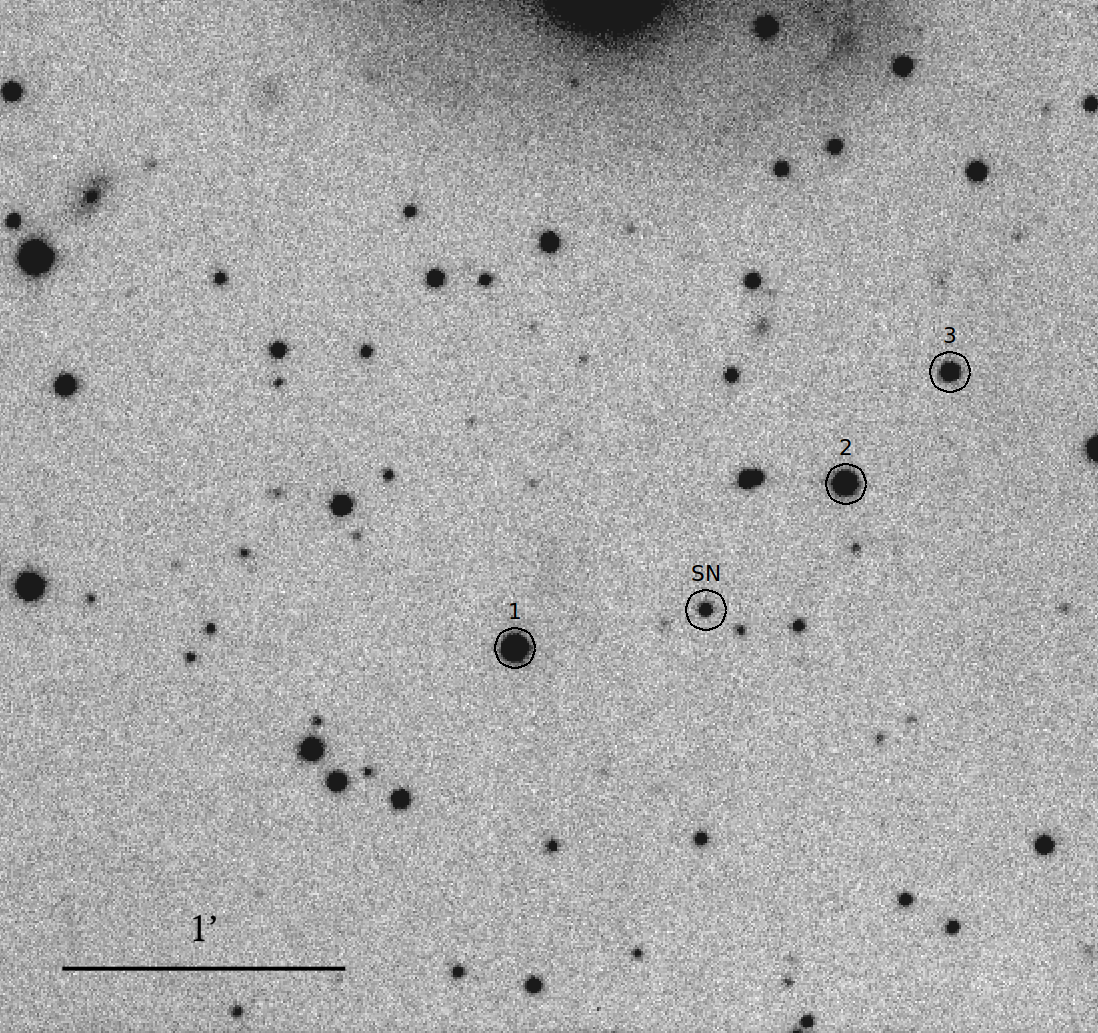}
    \caption{SN~2017gpn and comparison stars. The image is obtained with the Zeiss-1000 telescope in $R$ passband.}
    \label{fig:sn_frame}
\end{figure}

SN~2017gpn is located $\sim$140~arcsec from the centre of the host galaxy NGC~1343 (Fig.~\ref{fig:sn_frame}). Taking into account that the redshift of NGC~1343 is 0.0073~\citep{2005ApJS..160..149S} and assuming flat $\Lambda$CDM cosmology with $\Omega_\Lambda = 0.7$ and $H_0 = 70$~km~s$^{-1}$~Mpc$^{-1}$, we find that the \secrev{projected} distance between SN~2017gpn and the centre of its host is $\sim$21~kpc.

\subsection{Photometric data processing}
We performed 20 epochs of observations ($B$ and $R$ passbands) with the CCD photometer on the Zeiss-1000 telescope at SAO RAS. The aperture photometry was performed using standard procedures of the ESO~\textsc{MIDAS} software package. It includes standard image processing such as bias subtraction and flat field correction, removing the traces of cosmic particles, and stacking of individual frames into a summary image. 

Since no Landolt or any other standards~\citep{1987Stetson, 1992Landolt} were available for this region, we use the Pan-STARRS~\citep{2016Pan-starrs1, 2016Pan-starrs2} magnitudes for comparison stars. These magnitudes were \elena{recalculated} from the $g$, $r$, $i$ passbands to $B$ and $R$ with the use of Lupton's transformation equations\footnote{\url{http://www.sdss3.org/dr8/algorithms/sdssUBVRITransform.php}}: 

\begin{table}
    \centering
    \caption{Magnitudes of the comparison stars in the $B$ and $R$ passbands derived from $g$, $r$, $i$ Pan-STARRS1 magnitudes using Lupton's 2005 transformation equations.}
    \label{tab:comp_stars}
    \scalebox{1.0}{\begin{tabular}{lcccccr}
    \hline
        \textnumero & $B$ & err$_B$ & $R$ & err$_R$ \\
        \hline
        1 & 16.447  & 0.011      & 15.032  & 0.015 \\
        2 & 16.859  & 0.012      & 15.428  & 0.015 \\
        3 & 17.705  & 0.011      & 16.636  & 0.017 \\
        \hline
    \end{tabular}}   
\end{table}

\begin{equation}
      \begin{aligned}
    B &= g + 0.3130 \ (g - r) + 0.2271, \ &\sigma = 0.0107\\
    R &= r - 0.1837 \ (g - r) - 0.0971, \ &\sigma = 0.0106\\
    R &= r - 0.2936 \ (r - i) - 0.1439, \ &\sigma = 0.0072\\
      \end{aligned}
    \label{lupton}
\end{equation}
The comparison stars are shown in Fig.~\ref{fig:sn_frame} and their magnitudes are listed in Table~\ref{tab:comp_stars}.

We use a line-of-sight reddening for our Galaxy of $E(B - V) = 0.30$ mag~\citep{2011ApJ...737..103S}, corresponding to additive magnitude corrections of 1.246 and 0.725 mag for the $B$ and $R$ passbands, respectively. 
Since SN~2017gpn is very far from the centre of NGC~1343, we assume that the host's contamination is negligible. The resulting photometric data are presented in Table~\ref{tab:photometry}.

\subsection{Resulting light curves}
\begin{figure}
    \centering
	\includegraphics[width=1\columnwidth]{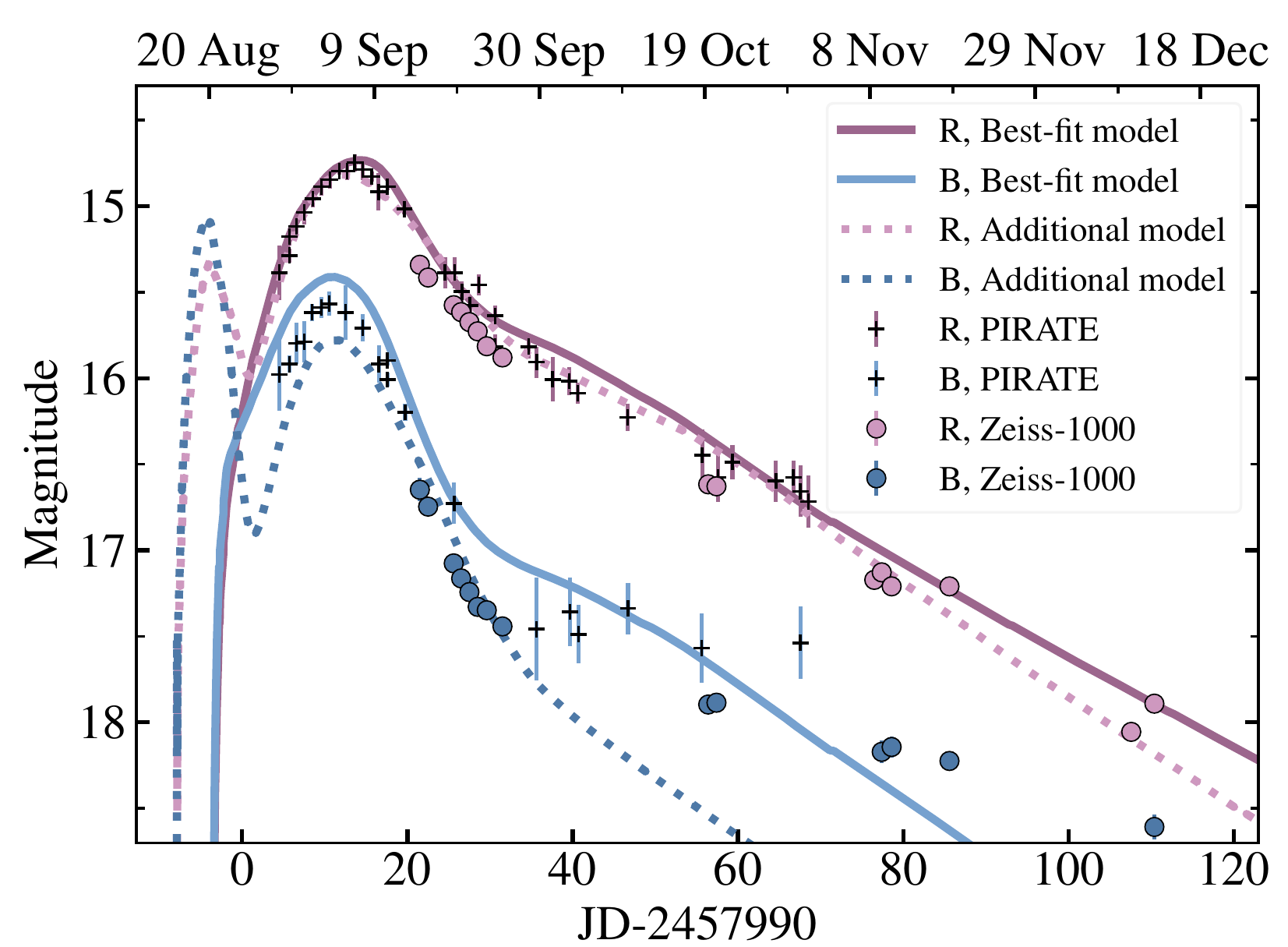}
    \caption{Light curve of SN~2017gpn. Pink and blue solid lines correspond to the best-fitting model; dashed lines to \maria{the} additional model in the $R$ and $B$ passbands, respectively. Circles are the Zeiss-1000 data; crosses are the data taken from~\citealt{2018Roberts}.}
    \label{fig:lightcurve}
\end{figure}

With Zeiss-1000 observations we can restore only the post-maximum part of the light curve. This is why, to improve the accuracy of \maria{the} further hydrodynamic modelling (see Section~\ref{sec:modelling}), we supplemented our data with observations in the $B$ and $R$ passbands from~\citealt{2018Roberts} obtained with the PIRATE robotic telescope in Spain~\citep{Holmes_2011}. The resulting light curve is presented in Fig.~\ref{fig:lightcurve}. The data points obtained at Zeiss-1000 (shown as circles) and the data points taken from~\citealt{2018Roberts} (marked with crosses) mutually complement each other and allow us to restore the $B$ and $R$ light curves almost entirely. 

One can notice a slight shift between the two data sets. This may be due to the different sources of photometry for the comparison stars since there are no Stetson and Landolt photometric standards in this field. However, the difference between the values is less than the uncertainty associated with the choice of hydrodynamic model; therefore for our purpose it can be neglected.

\begin{table}
	\centering
	\caption{Photometric observations of SN~2017gpn with \maria{the} Zeiss-1000 telescope. The magnitudes are corrected for the expected Galactic foreground extinction.}
	\label{tab:photometry}
	\scalebox{1.0}{\begin{tabular}{lccccr} 
		\hline
		\elena{JD 2457990+} & $B$ & err$_B$ & $R$ & err$_R$ \\
		\hline		
        21.5 & 16.65 & 0.07 & 15.34 & 0.03 \\
        22.5 & 16.75 & 0.05 & 15.41 & 0.02 \\
        25.6 & 17.08 & 0.05 & 15.58 & 0.03 \\
        26.5 & 17.16 & 0.05 & 15.62 & 0.04 \\
        27.5 & 17.24 & 0.06 & 15.67 & 0.02 \\
        28.5 & 17.33 & 0.06 & 15.73 & 0.02 \\
        29.6 & 17.35 & 0.05 & 15.81 & 0.02 \\
        31.5 & 17.44 & 0.06 & 15.88 & 0.03 \\
        56.4 & 17.90 & 0.06 & 16.62 & 0.02 \\
        57.4 & 17.89 & 0.05 & 16.63 & 0.01 \\
        76.5 & ---   & ---  & 17.17 & 0.03 \\
        77.4 & 18.17 & 0.07 & 17.13 & 0.03 \\
        78.6 & 18.14 & 0.06 & 17.21 & 0.03 \\
        85.6 & 18.22 & 0.05 & 17.21 & 0.03 \\
        107.6 & ---  & ---  & 18.06 & 0.04 \\
        110.4 & 18.61 & 0.07 & 17.89 & 0.03 \\
        143.3 & 19.14 & 0.15 & 18.78 & 0.01 \\
        153.3 & --- & --- & 18.54 & 0.30 \\
        224.3 & --- & --- & 21.14 & 0.20 \\		
		\hline
	\end{tabular}}
\end{table}

\subsection{Spectra}
\label{subsec:spectra}
The spectroscopic observations were collected using the Xinglong 2.16-m telescope and \maria{the} BFOSC system. All the spectra were reduced using routine tasks within IRAF and the flux was calibrated with spectrophotometric standard stars observed on the same nights. Telluric lines are removed from all of these spectra. The journal of our spectroscopic observations is given in Table~\ref{tab:spec_obs}. 

\begin{table*}
    \centering
    \caption{Journal of spectroscopic observations of SN~2017gpn with \maria{the} BFOSC+G4 instrument of the Xinglong 2.16-m telescope. Values of \maria{the} ejecta velocity measured from \maria{the} absorption lines of $\rm H~\alpha$\secrev{, \ion{He}{I}~$\lambda$5876, \ion{Fe}{II}~$\lambda$5018, and \ion{Fe}{II}~$\lambda$5169} are also presented.}
    \label{tab:spec_obs}
        \begin{tabular}{lccccc}
        \hline
\multirow{2}{*}{JD 2457990+} & Exp.~Time &  $\rm H~\alpha$ & \ion{He}{I}~$\lambda$5876 & \ion{Fe}{II}~$\lambda$5018 & \ion{Fe}{II}~$\lambda$5169 \\
  & [s]  & [km~s$^{-1}$] & [km~s$^{-1}$] & [km~s$^{-1}$] & [km~s$^{-1}$]\\
\hline
5.30 & 3600 & 15000 $\pm$ 130 & 10100 $\pm$ 300 & 12000 $\pm$ 1200 & 11400 $\pm$ 950\\
25.29 & 3600 & 13200 $\pm$ 100 & 8000 $\pm$ 100 & 6750 $\pm$ 470 & 5130 $\pm$ 490\\
33.33 & 2700 & 12900 $\pm$ 200 & 7300 $\pm$ 200 & --- & --- \\
        \hline
        \end{tabular}
\end{table*}

\begin{figure}
    \centering
	\includegraphics[width=1\columnwidth]{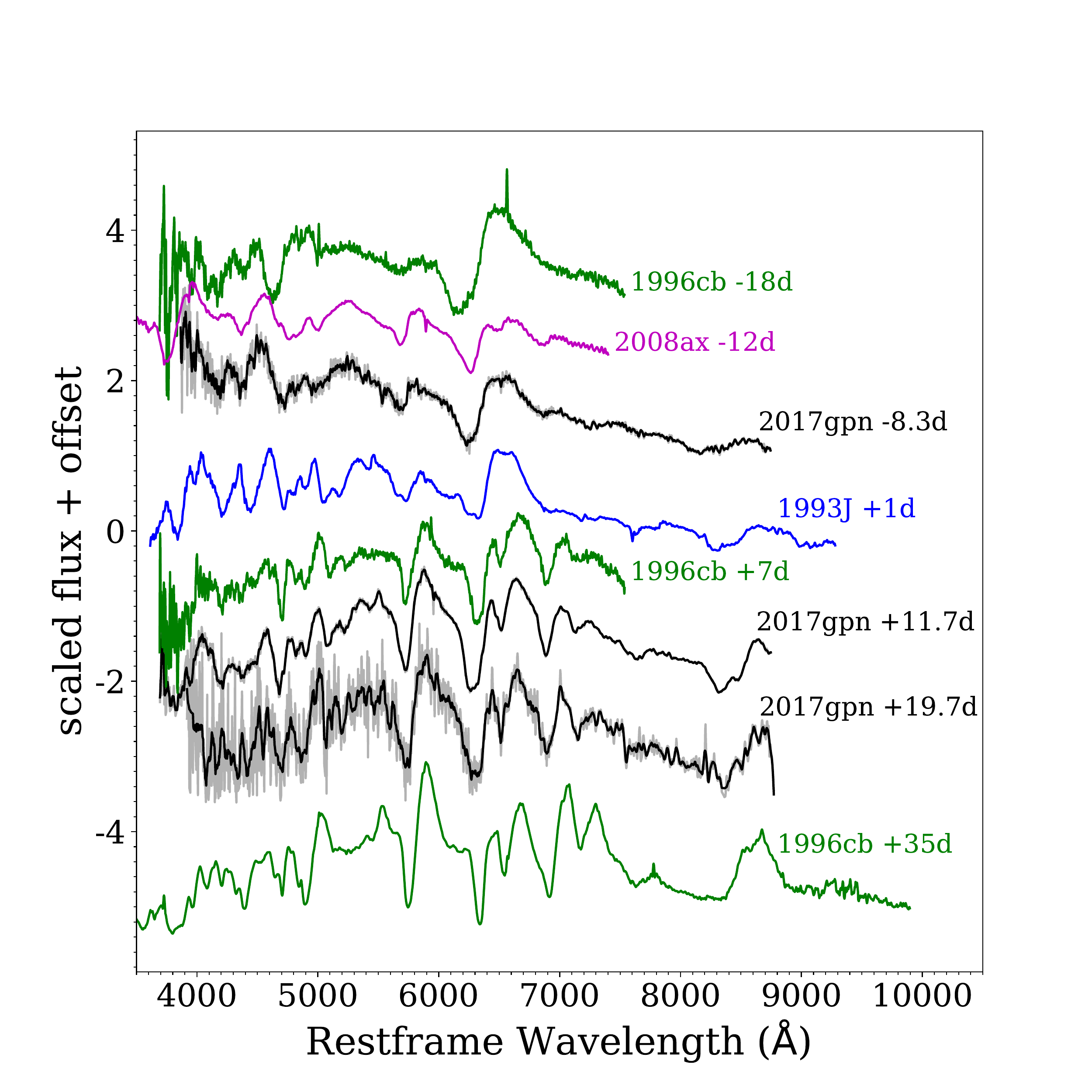}
    \caption{Three spectra of SN~2017gpn at different phases, the observation dates are indicated with respect to the \maria{$R$-band maximum light at JD$~=~2458003.6$}. Spectra of SNe~IIb 1993J, 1996cb, and 2008ax are presented for comparison.}
    \label{fig:spectra}
\end{figure}

Three optical spectra were obtained for SN~2017gpn, covering the phases \elena{from $-8.3$ to $+19.7$ d} from the $R$-band maximum light (peak time is JD$~=~2458003.6$); these are shown in Fig.~\ref{fig:spectra}. At one week before the peak, the spectrum shows strong Balmer lines of hydrogen, providing evidence of a Type~II Supernova. Moreover, the existing prominent absorption features at $\sim$5670 and 6860~\AA\AA~that can be identified as \ion{He}{I}~$\lambda$5876 and \ion{He}{I}~$\lambda$7065, respectively, confirming that SN~2017gpn can be further put into the Type~IIb subclass. From the absorption minima of $\rm H~\alpha$ and \ion{He}{I}~$\lambda$5876 lines at the first obtained spectrum, we measured the ejecta velocity as 15000~$\pm$~130 and 10100~$\pm$~300~km~s$^{-1}$, respectively, indicating that the Balmer lines and the \ion{He}{I} lines originated from different layers (see Table~\ref{tab:spec_obs}). At two weeks after the maximum, the helium features seem to become more noticeable and other helium features such as \ion{He}{I}~$\lambda$6678 (blueshifted to $\sim$6510~\AA) emerge in the spectrum. The helium features become even more pronounced in the spectrum taken one week later, while the hydrogen features become gradually weaker. The overall spectral evolution of SN~2017gpn is presented in Fig.~\ref{fig:spectra} and it is similar to other typical Type~IIb Supernovae, like SN~1993J~\citep{1995A&AS..110..513B}, SN~1996cb~\citep{1999AJ....117..736Q}, and SN~2008ax~\citep{2014AJ....147...99M}. 

\section{Modelling}
\label{sec:modelling}
\maria{\subsection{Pre-supernova models}
A set of non-evolutionary pre-supernova models is obtained under the assumption of a power-law dependence of temperature on density: $T\propto\rho^\alpha$~\citep{1986NInfo..61...29N,1993A&A...273..106B}. Therefore, the obtained hydrostatic configuration would be close to a polytrope of index $1/\alpha \simeq 3$. The deviation from the polytropic model increases in the outer layers due to recombination of ions and non-homogeneous chemical composition.}

\maria{At the centre of such a configuration we isolated a point-like source of gravity that has a non-negligible influence on the expansion of the innermost layers of supernova ejecta. The mass and radius of this compact remnant are taken as $\rm M_{CR}=1.41$~$M_\odot$ and 0.01~$R_\odot$ for all treated pre-SN models.}

\maria{In our approach we do not follow the explosive nucleosynthesis. Thus, the SN ejecta composition is the same as the pre-SN composition except for $^{56}\rm Ni$. Since the amount and distribution of $^{56}\rm Ni$ synthesized during the explosion plays a crucial role in the SN luminosity evolution, we consider two radial distributions for $^{56}\rm Ni$. In the first one $^{56}\rm Ni$ is totally mixed through the ejecta and in the second one $^{56}\rm Ni$ falls off from the centre.}

As input parameters for further hydrodynamical modelling, we varied the pre-SN star mass $\rm M$ and \maria{the} radius $\rm R$, the mass of synthesized nickel $\rm M_{^{56}Ni}$, and the initial distribution of chemical elements in the pre-SN star.

\subsection{\textsc{STELLA} code}
\maria{To explode the hydrostatic non-evolutionary pre-SN models a one-dimensional multifrequency radiation hydrocode \textsc{STELLA} is used.} The full description of the code can be found in~\cite{Blinnikov1998,Blinnikov2006}; a public version of \textsc{STELLA} is also included with the \textsc{MESA} distribution~\citep{Paxton2018}. The \textsc{STELLA} code is used for the light-curve modelling of different types of SNe --- Ia~\citep{Blinnikov2006}, Ib/Ic~\citep{Folatelli2006,Tauris2013}, IIb~\citep{Blinnikov1998,Tsvetkov2012}, IIn~\citep{Chugai2004}, IIP~\citep{Baklanov2005,Tominaga2009}, Ic associated with long gamma-ray bursts~\citep{2017MNRAS.467.3500V}. \maria{The} \textsc{STELLA} code was compared with other well-known hydrodynamic codes and found to be in good agreement with them on the level of several per cent (e.g.~\citealt{2007ApJ...662..487W,2009MNRAS.398.1809K,2010ASPC..429..148S,2017MNRAS.464.2854K,2020arXiv200601832T}). 

In the current calculations, we adopted 100 zones for the Lagrangian coordinate and 130 frequency bins. \maria{The explosion is initiated by putting thermal energy into the innermost layers. The energy is released in 0.1~s, which is less than the hydrodynamic time of the pre-supernova. While this condition is true, the resulting light curve is not affected by the details of the explosion mechanism~\citep{1983ASPRv...2...75I}.}

\subsection{Best-fitting model}

To determine the best-fitting model of SN~2017gpn we consider a grid of parameters. The pre-SN mass varies between $3.5 \ M_\odot$ and $5.5 \ M_\odot$ with steps of $0.5 \ M_\odot$; the pre-SN radius and $\rm E_{exp}$ take the values $\{50, 100, 200, 400, 600\} \ R_\odot$ and $\{0.6, 1.2, 2.4\}\times 10^{51}$~erg, respectively; three different $\rm M_{^{56}Ni}$ $\{0.07, 0.09, 0.11\} \ M_\odot$ are considered, both with and without mixing. The mass of the hydrogen envelope $\rm M_{H\_{env}}$ is taken as 0.06~$M_\odot$ which is in line with our expectations for Type~IIb Supernovae.

After determination of the parameter grid we built trial models and chose the best-fitting model within the generated grids of light curves by calculating $\chi^2$ in the $R$ passband. The best-fitting model corresponds to the minimum value of $\chi^2$. \maria{We do not provide any statistical uncertainties, since this procedure requires enormous computational effort. Instead, the optimal model is recovered as a compromise between the fits to the observed light curve and the evolution of the velocity at the photosphere (see Section~\ref{ph_vel}).} The values of \maria{the} best-fitting model parameters are summarized in Table~\ref{tab:result}. Fig.~\ref{fig:lightcurve} compares the light curves of the model (solid lines) with the observations of SN~2017gpn. 

\begin{table}
    \caption{Parameters for the best-fitting and additional hydrodynamic models of SN~2017gpn.}
    \label{tab:result}
    \centering
    \scalebox{1.0}
    {
    \begin{tabular}{lcr}
    \hline
Parameter & Best-fitting model  & Additional model \\
\hline
$\rm R$  & $50~R_{\odot}$  &  $400~R_{\odot}$ \\
$\rm M$  & $3.5~M_{\odot}$  & $3.5~M_{\odot}$  \\
$\rm M_{H\_{env}}$ & $0.06~M_{\odot}$  & $0.21~M_{\odot}$ \\
$\rm M_{CR}$ & $1.41~M_{\odot}$  & $1.41~M_{\odot}$ \\
$\rm M_{^{56}Ni}$ & $0.11~M_{\odot}$, mixed  & $0.11~M_{\odot}$, no mixing \\
$\rm E_{exp}$ & $1.2 \times 10^{51}$ erg  & $1.2 \times 10^{51}$ erg\\
$\rm t_{peak,{\textit R}}$ & 2017 Sept 7.5 & 2017 Sept 5.6\\
    \hline
    \end{tabular}
    }
\end{table}

\begin{figure}
    \centering
	\includegraphics[width=1\columnwidth]{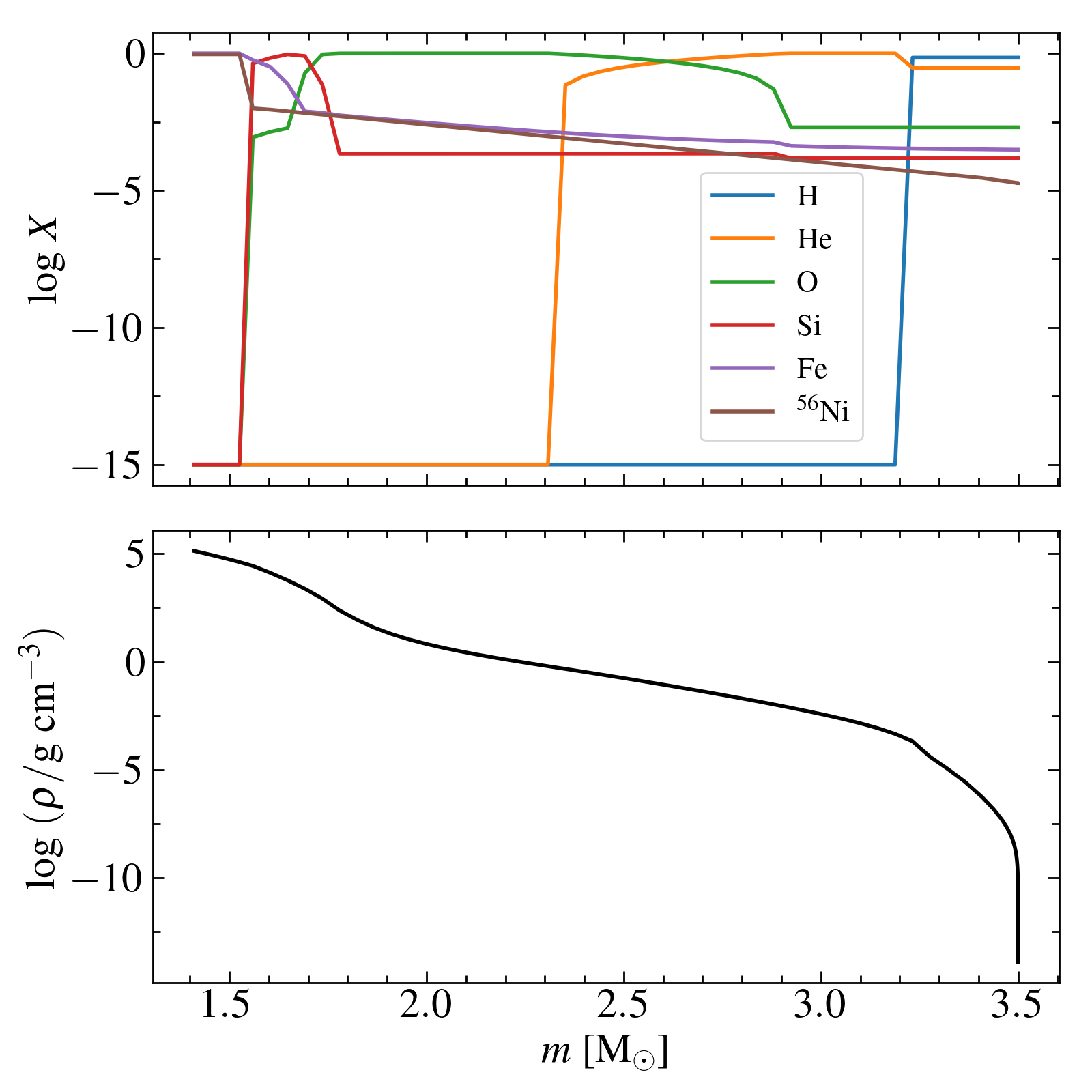}
    \caption{Mass fractions \maria{of the most abundant} chemical elements in the ejecta (top) and density profile (bottom) for the best-fitting pre-SN star model with respect to the interior mass. The central region of 1.41~$M_\odot$ is taken away.}
    \label{fig:rho}
\end{figure}

In Fig.~\ref{fig:rho} we also show the distribution of the chemical elements and the density profile for a pre-SN star. Note that the best-fitting model shows a small amount of hydrogen in the pre-SN star composition, which is expected for SNe~IIb~\citep{1993Filippenko}. $\rm ^{56}Ni$ is totally mixed through the ejecta.

\subsection{The influence of the model parameters on the light curve}
\label{subsec:influence}
\begin{figure*}
\begin{center}
\includegraphics[width=2\columnwidth]{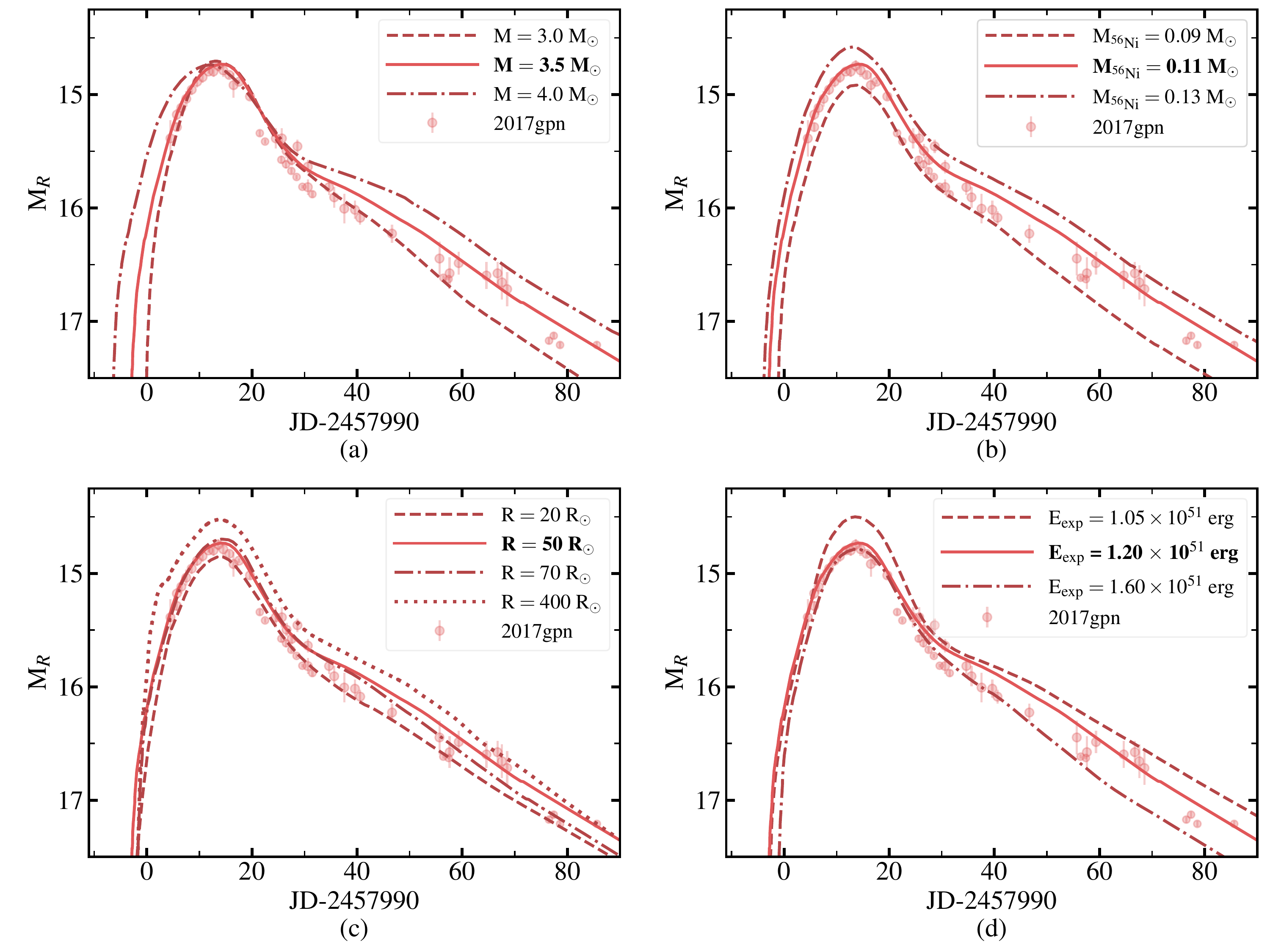}
\caption{\secrev{The dependence of the modelled $R$-passband LC on the pre-SN mass~M~(a), the amount of synthesized $\rm ^{56}Ni$~(b), the pre-SN radius R~(c), and the explosion energy~$E_{exp}$~(d). All models are shifted along the time axis to better describe the observations. The best-fitting model is shown by a solid line in all plots; observations are shown by circles.}}
\label{fig:Rband}
\end{center}
\end{figure*}

\secrev{To provide a reasonable range of the best-fitting model parameters, we consider the dependence of the numerical LCs on an input parameter of the model while the others remain fixed. We vary the mass~$\rm M$ and the radius~$\rm R$ of the pre-SN star, the mass of synthesized $\rm ^{56}Ni$, and the energy of the explosion~$\rm E_{exp}$. In Fig.~\ref{fig:Rband} we plot some modelled LCs in the $R$~passband that show a valid range for each parameter. All presented models are slightly shifted along the time axis to better describe the observational light curve.}

\secrev{The first considered parameter is the pre-SN mass~$\rm M$, see Fig.~\ref{fig:Rband}~(a). This parameter mainly affects the width of the light curve, which becomes broader as the mass increases. This is explained by the fact that with a small mass the envelope becomes transparent faster. Thus, the LC increases before the maximum light and decreases rapidly after it. As Fig.~\ref{fig:Rband}~(a) shows, the range of valid pre-SN mass is 3--4~$M_\odot$.} 

\secrev{The next parameter is the amount of synthesized ${\rm ^{56}Ni}$ (Fig.~\ref{fig:Rband}~(b)). The models are brighter for higher ${\rm ^{56}Ni}$ masses. The LCs corresponding to the ${\rm ^{56}Ni}$ masses of $0.09$ and $0.13~M_\odot$ lie below and above the best-fitting model light curve, respectively. These two values define the acceptance range of the ${\rm M_{^{56}Ni}}$ model parameter.}

\secrev{The pre-SN radius affects mainly the light-curve tail: a larger radius value corresponds to a brighter light curve after maximum light. The chosen range of the pre-SN radius is 20--70~$R_\odot$; see Fig.~\ref{fig:Rband}~(c).}

\secrev{The last parameter that we vary is the explosion energy~$\rm E_{exp}$; see Fig.~\ref{fig:Rband}~(d). The determined range for the energy parameter is (1.05--1.60)$\times 10^{51}$~erg. As seen from Fig~\ref{fig:Rband}~(d), smaller values of $\rm E_{exp}$ correspond to brighter light curves. This dependence is in line with our expectations. A larger $\rm E_{exp}$, for a fixed mass of $^{56}$Ni and fixed total mass, implies higher velocities and hence less trapping of gamma-ray photons. This leads to an increase in the predicted observed gamma-ray flux and, therefore, to a decrease in the emission in the visible light range.}

\section{Discussion}
\label{sec:discussion}
\subsection{Comparison with other SNe~IIb}
\begin{figure*}
\begin{center}
\includegraphics[width=2\columnwidth]{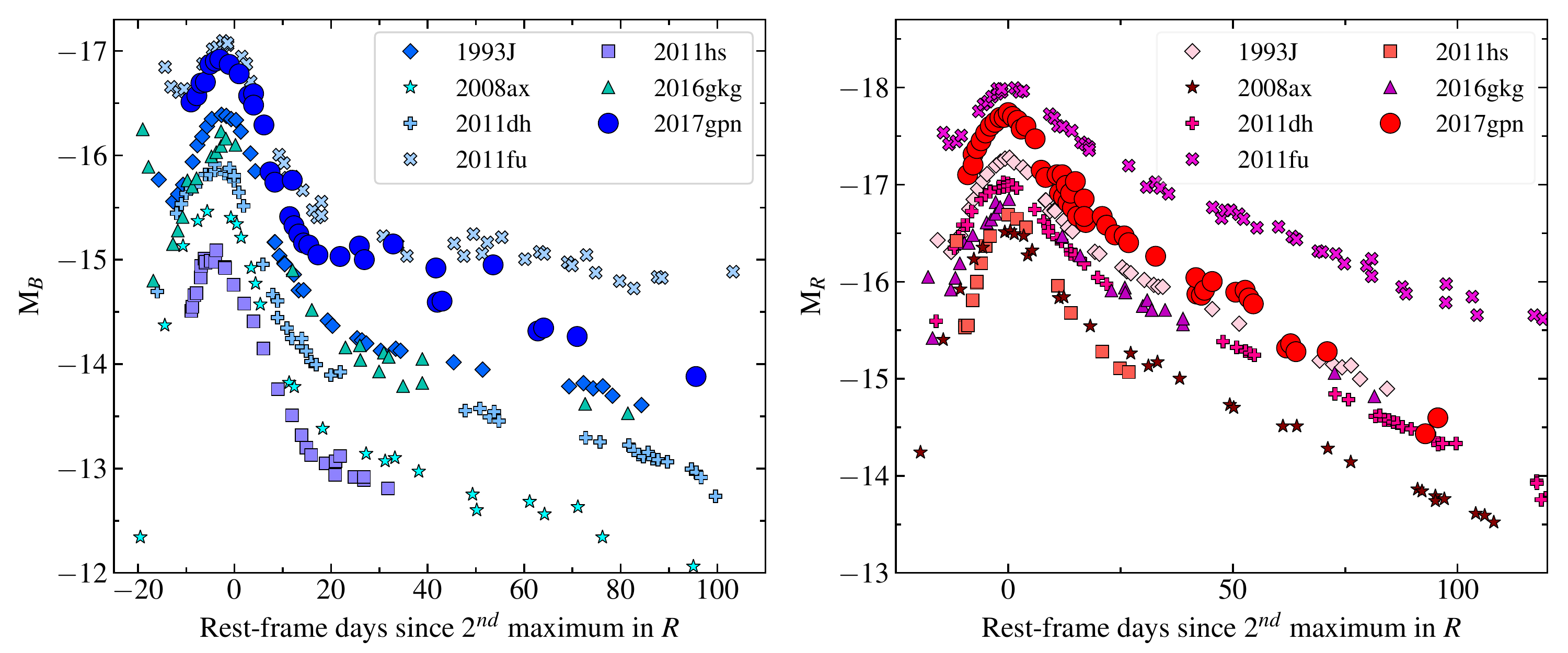}
\caption{M$_B$ and M$_R$ light curves of SN~2017gpn in comparison with those of other Type~IIb Supernovae: 1993J~\citep{1996AJ....112..732R}, 2008ax~\citep{2009PZ.....29....2T}, 2011dh~\citep{Tsvetkov2012}, 2011fu~\citep{2013MNRAS.431..308K}, 2011hs~\citep{2014MNRAS.439.1807B}, 2016gkg~\citep{2018Natur.554..497B}.}
\label{fig:lcIIb}
\end{center}
\end{figure*}

\elena{We collected data for well-studied SNe~IIb with good photometric coverage in the $B$ and $R$ passbands, for which results of hydrodynamic modelling can be found in the literature. 
In Fig.~\ref{fig:lcIIb} the light curves of selected SNe~IIb are presented.} It can be seen that LCs in the $B$ and $R$ passbands are similar --- characteristic bell-shaped LCs. Moreover, as~\cite{2019Pessi} showed, SNe~IIb take longer to reach maximum light and decline more quickly post-maximum than hydrogen-rich SNe~II, so the authors assume that there is no continuum \elena{between SNe~IIb and other SNe~II} like between SNe~IIP and IIL~types. SN~2017gpn has a typical SN~IIb light curve, and belongs to one of the brightest well-studied SNe~IIb: it is brighter than a typical member of Type~IIb SN~1993J by 0.75~mag in the $R$ passband.

\subsubsection{Classification of \citet{PM2017}}
\begin{figure}
    \centering
	\includegraphics[width=1\columnwidth]{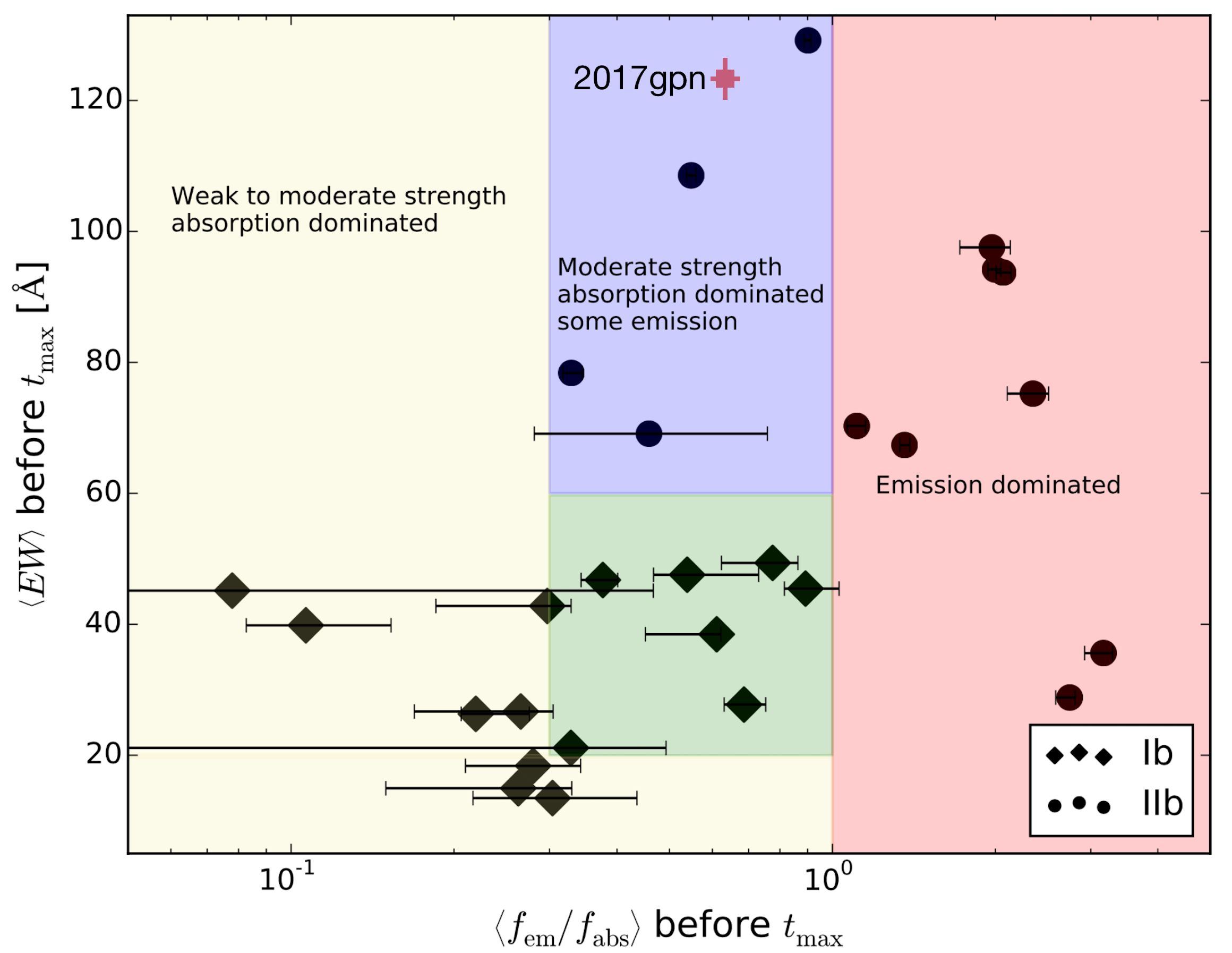}
    \caption{\elena{This figure is borrowed from~\citealt{PM2017} (fig. 7) with SN~2017gpn plotted (pink square). It illustrates the stripped-envelope supernovae subclassification based on the comparison of the line strength (equivalent width of $\rm H~\alpha$) against the line profile ($\rm H~\alpha$~emission-to-absorption ratio $f_{em}/f_{abs}$) as proposed by~\citet{PM2017}. SN~2017gpn lies in the blue region that corresponds to group~IIb(I). Groups IIb, Ib, and Ib(II) are in the red, yellow, and green regions, respectively.}}
    \label{fig:sub-class}
\end{figure}

\elena{Following~\citet{PM2017}, stripped-envelope SNe should be subclassified into four groups: Ib, Ib(II), IIb, and IIb(I), using the additional parameters --- equivalent width of $\rm H~\alpha$ ($EW_{\rm H~\alpha}$) and $\rm H~\alpha$~emission-to-absorption ratio $f_{em}/f_{abs}$. The $EW_{\rm H~\alpha}$ parameter value is $>60$\AA~for supernovae of group IIb(I), $20<EW_{\rm H~\alpha}<60$\AA~for Ib(II), and takes any reasonable value for groups Ib and IIb. The $\rm H~\alpha$~emission-to-absorption ratio $f_{em}/f_{abs}$ differs for groups IIb and IIb(I): it ranges from 0.3 --- 1 for IIb(I) and is greater than 1 for group~IIb (see Fig.~\ref{fig:sub-class}).}

\elena{We calculated the intensity and equivalent width of $\rm H~\alpha$ in our first spectrum ($-8.3$ d before $R$-band maximum) for SN~2017gpn and found $f_{em}/f_{abs}= 0.63\pm 0.04$, $EW_{\rm H~\alpha}=123\pm3$\AA. Therefore, SN~2017gpn belongs to group IIb(I) which means that it might have less hydrogen in the envelope than most $\rm H$-rich SNe such as 1993J, 2011fu, or 2011dh (see Table~\ref{tab:othersn2b}). However, it is similar to other SNe~IIb(I) --- 2008ax and 1996cb (the first position in a cross-correlation list according to \textsc{SNID}).}

\subsubsection{Hydrodynamic models of other SNe~IIb}
\elena{We compare the results of numerical simulations for SN~2017gpn and other SNe~IIb (including groups IIb and IIb(I) of~\citealt{PM2017}) presented in Fig.~\ref{fig:lcIIb}. Only hydrodynamic modelling of supernovae is chosen for comparison; we do not consider any analytical light-curve modelling or scaling to templates. The modelling results are summarised in Table~\ref{tab:othersn2b}, where $\rm M_{CR}$ is the mass of a compact object (generally this is a neutron star) and M$\rm_{ej}$ is the mass of ejected matter.}

\elena{The main modelling parameters such as the ejecta mass $\rm M_{ej}$, the mass of $\rm ^{56}{Ni}$, $\rm M_{H\_env}$, and the explosion energy $\rm E_{exp}$ are consistent with each other. An exception is the parameter of the pre-supernova radius $\rm R$. The considered hydrodynamic modelling shows that the pre-SN radius lies in a broad range from 30~---~720~$R_{\odot}$ and may be different for the same object in different models. For example, there are two models for SN~2008ax, one with a radius of 30--50~$R_{\odot}$~\citep{2015ApJ...811..147F} and another one with $\rm R=600$~$R_{\odot}$~\citep{2009PZ.....29....2T}. It should be noted that SN~2008ax belongs to the same group of IIb(I) supernovae as SN~2017gpn.}

\subsection{Additional model}
\label{subsec:add_model}
Motivated by the discrepancy in the modelled radius for different SNe~IIb, we have found another physically reasonable model for SN 2017gpn with $\rm R = 400$ $R_{\odot}$. For this additional model, radioactive nickel is located in the central part of the ejecta. We have also increased the mass of \maria{the} hydrogen envelope to $0.21$~$M_\odot$, \maria{which is consistent with the fact that more extended SNe~IIb should be also more $\rm H$-rich~\citep{PM2017}.} The parameters of the additional model are listed in Table~\ref{tab:result}. This model also well describes the observational data and agrees with the results of \maria{the} hydrodynamic simulations for other SNe~IIb. 

\maria{There is no direct method to solve the inverse problem, i.e. to determine the parameters of the pre-supernova from the observational data. We can only build a model with given parameters and see how accurately it fits the data. Sometimes it can happen that models with different parameters reproduce observations equally well, as we see for our best-fitting and additional models (Fig.~\ref{fig:lightcurve}). However, if some additional information is available, e.g. observational photospheric velocities, we can compare our theoretical estimations with the observational values and make a choice between the models.}

\subsubsection{Photospheric velocities}
\label{ph_vel}
\begin{figure}
    \centering
        \includegraphics[width=0.95\columnwidth]{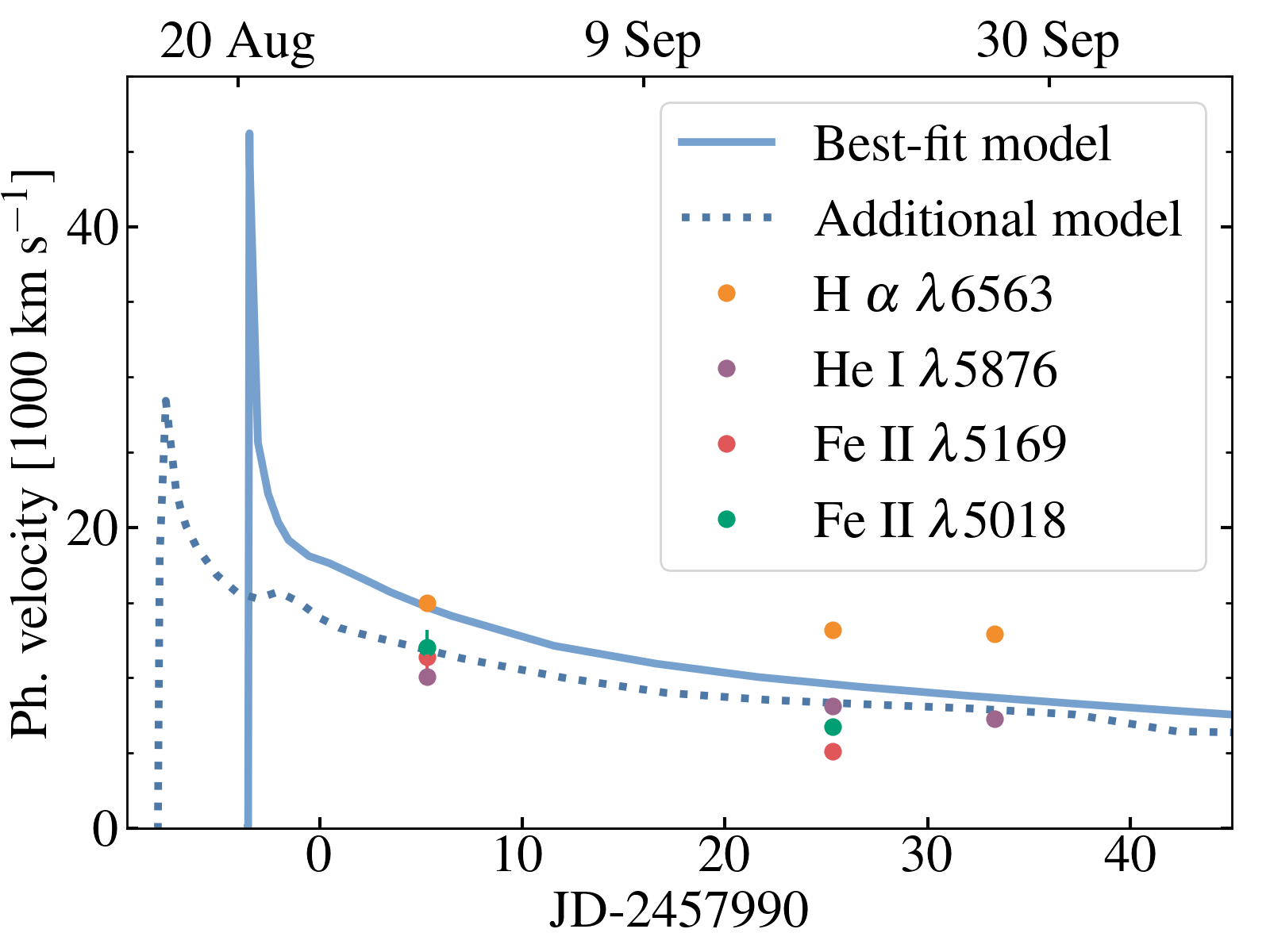}
        \caption{Photospheric velocity \maria{at the $\tau=2/3$ level as a function of} time for the best-fitting model (solid line) and for the additional model of higher radius (dashed line); dots are the observational velocities measured from the $\rm H~\alpha$, \ion{He}{I}~$\lambda$5876, \secrev{\ion{Fe}{II}~$\lambda$5018, and \ion{Fe}{II}~$\lambda$5169} absorption lines.}
    \label{fig:vph1}
\end{figure}

Based on three spectra of SN~2017gpn obtained at different epochs with the Xinglong 2.16-m telescope, we measured the ejecta velocity from the $\rm H~\alpha$ and \ion{He}{I}~$\lambda$5876 absorption lines (Table~\ref{tab:spec_obs}). In Fig.~\ref{fig:vph1} we show the comparison between the velocities measured from these lines and theoretical values from the \textsc{STELLA} code, which are the velocities of the photosphere at the $\tau=2/3$ level in the $B$ band. The best-fitting model is consistent with the velocity measured from the $\rm H~\alpha$ line for this epoch; the additional model is in good agreement with the \ion{He}{I}~$\lambda$5876 velocities for all three epochs. 

\secrev{It should be noted that P~Cygni profiles are formed in all layers above the photosphere. Hence, the hydrogen and helium features do not necessarily reflect the photospheric velocities calculated by our hydrodynamic modelling. It has to be taken into account that the growth of the Sobolev optical depth~\citep{1960Sobolev} at the photosphere level causes a significant blueshift of the P~Cygni profile minimum, so the resulting velocity in that case will be overestimated~\citep{2002Kasen}. This effect may explain why the velocities measured from the $\rm H~\alpha$ line are greater than our theoretical estimates in Fig.~\ref{fig:vph1}. Meanwhile, according to \citealt{2005Dessart, 2006Dessart} the velocities measured from strong lines can be both smaller and larger than the photospheric ones.}

\secrev{Therefore, it is more reasonable to use ``weak'' lines, i.e. lines with small Sobolev optical depths, to estimate $v_{ph}$. \citep{2005Dessart} show that \ion{Na}{I~D}, \ion{Fe}{II}~$\lambda$5018, \ion{Fe}{II}~$\lambda$5169 are the most suitable lines to measure the photospheric velocities. We measured the velocities from the \ion{Fe}{II}~$\lambda\lambda$5018~and~5169 lines for the first and second epochs of observations, the last epoch spectrum has a low signal-to-noise ratio to perform the measurements. We could not determine the velocities using \ion{Na}{I~D} features since they are close to \ion{He}{I}~$\lambda$5876 line, which is quite strong in SNe~IIb.}

\secrev{The photospheric velocities derived for the additional model correspond slightly better to the velocities from the \ion{Fe}{II}~$\lambda\lambda$5018~and~5169 lines for the first epoch of observations. For the second epoch the measured velocities are lower than the \textsc{STELLA} values for both models. Taking into account the modelling uncertainties, it is difficult to choose between the models based on these measurements only.}

\subsubsection{$^{56}$Ni mixing}
\begin{figure}
    \centering
        \includegraphics[width=0.95\columnwidth]{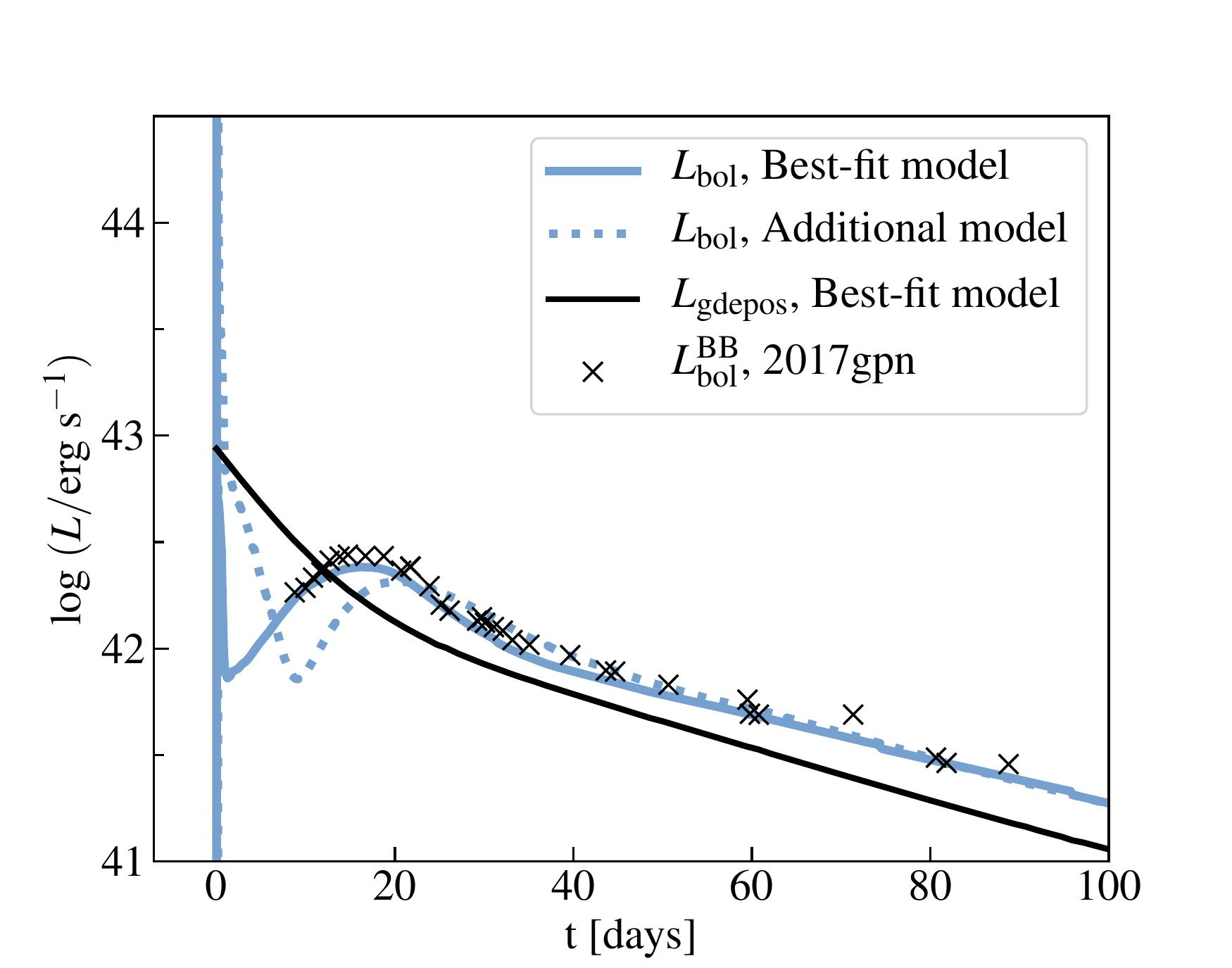}
        \caption{Theoretical bolometric light curves for the best-fitting (blue solid line) and the additional (blue dashed line) models of SN2017gpn. \secrev{The crosses show the bolometric luminosity of SN~2017gpn calculated from $B$ and $R$ light curves with use of the \textsc{SuperBol} code~\citep{Nicholl2018}. The shift between the data and the best-fitting model is the same as in Fig.~\ref{fig:lightcurve} but transformed to the rest frame.} The black solid line is the power due to the gamma-ray deposition from $^{56}$Ni and $^{56}$Co decays for our best-fitting model. Accounting for the light travel time correction, $L_{\rm gdepos}$ satisfies Arnett's law, going through the maximum of $L_{\rm bol}$.}
    \label{fig:gdepos}
\end{figure}

From the theoretical bolometric LCs (Fig.~\ref{fig:gdepos}) as well as LCs in filters (Fig.~\ref{fig:lightcurve}) it can be noticed that the light curve corresponding to the model with the uniform distribution of nickel behaves differently from the light curve that conforms with the model where nickel is concentrated in the centre of the ejecta. 
This is due to the fact that in the former case the radioactive decay energy contributes to the overall energy immediately after the explosion, whereas in the latter case we observe two peaks in the light curve. The primary peak is associated with the heating of the outer layers of the star by the shock wave that is created by the rebound of the freely falling inner layers from the collapsed core. After that the envelope expands, cools, and therefore becomes transparent. The second peak is associated with the luminescence of the inner layers heated by the radioactive decays of $^{56}$Ni and its products. For the additional model we fit the observed LCs by the second peak. Because of this, the best-fitting and additional models are shifted relative to each other in Fig.~\ref{fig:lightcurve}. \secrev{The influence of $^{56}$Ni mixing on the LC~behaviour is also seen if we compare the additional model with the model in Fig~\ref{fig:Rband}~(c) (dotted line) with $\rm R=400~R_\odot$ and $^{56}$Ni totally mixed through the ejecta. Unlike the additional model, this model no longer describes the observations.}

\secrev{In Fig.~\ref{fig:gdepos} we also show the bolometric light curve of SN~2017gpn restored from the available photometry. To construct the bolometric light curve the \textsc{SuperBol} code is used~\citep{Nicholl2018}. To account for flux that is not covered by the observations, the black body extrapolation is applied. Even though we use only two passbands ($B$ and $R$) the obtained bolometric LC agrees very well with our theoretical estimations.}

\begin{table*}
	\centering
	\caption{Comparison of \maria{the} hydrodynamic modelling results for different SNe~IIb.} 
	\label{tab:othersn2b}
	\scalebox{0.98}{\begin{tabular}{lccccccr}
\hline
SN name & $\rm M_{CR}$  & $\rm M_{ej}$ & $\rm M_{^{56}{Ni}}$  &  $\rm M_{H\_env}$ & $\rm R$   & $\rm E_{exp}$  & Reference \\
	    & [M$_\odot$]   & [M$_\odot$] & [M$_\odot$]         &  [M$_\odot$]  & [R$_\odot$] & [10$^{51}$ erg]   & \\		 
\hline		
\multirow{ 2}{*}{1993J} &\multirow{ 2}{*}{$\sim$1.4} & \multirow{ 2}{*}{1.4--3.1} & \multirow{ 2}{*}{0.06--0.08} & \multirow{ 2}{*}{0.2}& \multirow{ 2}{*}{430--720} & \multirow{ 2}{*}{1.2} & \cite{1994Woosley} \\
                 &&&&&&        & \cite{Blinnikov1998} \\	
\hline	
2008ax  & 1.41 &2.39 &0.11& -- &600 & 1.5  &~\cite{2009PZ.....29....2T} \\		
2008ax  & 1.5   & 1.8--3.5 &0.05-0.07& 0.06 &30--50 & 0.8--1.2 &~\cite{2015ApJ...811..147F} \\	
\hline
2011hs  &  1.5 & 1.5--2.5 &0.04& <0.5& 500--600&0.85  &~\cite{2014MNRAS.439.1807B} \\			
\hline
2011fu   & 1.5 & 3.5 & 0.15 &0.3 &450 &1.3 &~\cite{2015MNRAS.454...95M} \\	
\hline	
2011dh  & 1.41 & 2.24--4.24&0.07& --&150--300&2--4 & \cite{Tsvetkov2012}\\
2011dh  &  1.5 & 2 & 0.06 & 0.1 & 200 & 0.6--1  & \cite{2012ApJ...757...31B}\\
\hline	
2016gkg & 1.4 & 3.55   & 0.2  & 0.02  & 180--260 & 1.3  & \cite{2017ApJ...846...94P}\\
2016gkg & 1.5--1.6 & 2.5--3.4 & 0.085--0.087 & 0.01--0.09 &300-340 & 1--1.2 & \cite{2018Natur.554..497B}\\
		\hline
	\end{tabular}}
\end{table*}

\subsection{Arnett's law}
\maria{Arnett's law~\citep{Arnett1982} states that the energy released on the surface at maximum light is equal to the energy deposed by gamma-ray radiation. This law is commonly used to estimate the amount of nickel produced in the explosion when the total luminosity at peak is known~\citep{Branch1992}. We plot the theoretical bolometric light curve and the curve corresponding to gamma-ray deposition from $^{56}$Ni and $^{56}$Co decays for our best-fitting model to check this law. As we can see from Fig.~\ref{fig:gdepos}, the law is quite well satisfied; however, the power from gamma-ray deposition does not go directly through the $L_{\rm bol}$ peak. This is explained by the fact that Arnett's law is not exact and in particular assumes an infinite speed of light. In the \textsc{STELLA} code the energy released in the centre will be ``seen'' with a delay of $R/c$, where $R$ is the radius of the expanding ejecta that changes with time and $c$ is the speed of light. The observed difference increases towards the tail since the radius increases as well.}

\subsection{SN~2017gpn position relative to the host galaxy centre}
\label{subsec:separation}
Supernova 2017gpn exploded in the \secrev{spiral} galaxy NGC~1343 at a \secrev{projected} distance of $D\simeq21$~kpc from its centre (see Fig.~\ref{fig:sn_frame}). 
Such location is unusual for core-collapse supernovae, in particular for Type~IIb, since it is believed that stripped-envelope CCSNe are formed from very massive stars in star-formation regions of galaxies (see~\citealt{2020MNRAS.492..848A} and references therein). \secrev{Assuming that SN~2017gpn belongs to the galactic disk we can take into account the projection effect. The deprojected distance $D_{\rm dep}$ between the supernova and the host centre is calculated as}

\begin{equation}
    D_{\rm dep} = D\sqrt{\cos^2\alpha + \sin^2\alpha \sec^2i},
\end{equation}
\secrev{where $\alpha$ is the angle between the projected distance and the major axis of a galaxy and $i$ is the disc inclination angle. According to HyperLEDA $i$~equals 67.3~deg and the major axis position angle of NGC~1343 is 78.8~deg~\citep{Hyperleda}. Using these values and the coordinates of SN~2017gpn and its host galaxy centre we can calculate the deprojected distance for SN~2017gpn, which is~$\sim$52~kpc.} To understand how exceptional this position is we study the absolute and relative separations between the supernova positions and their host galaxy centres for a sample of SNe~IIb. 

\secrev{Hereafter}, by the distance between a supernova and its host galaxy we mean the projection of the distance onto the picture plane, which is obviously smaller than the real distance.
However, the star-evolution theory predicts that CCSNe including SNe~IIb mainly appear in the galactic planes of spiral galaxies, in regions of high star-formation rate. Therefore, we assume that the contribution of the projection onto the line of sight is relatively small and this underestimation of the distance does not significantly affect our analysis.

We collected 71 confirmed SNe~IIb and 108 candidates for SNe~IIb from the Open Supernova Catalog~\citep{SneSpace}. The confirmed SNe~IIb are supernovae for which multiple spectra have been obtained and a detailed spectral analysis has been performed. If only \maria{a spectrum is} available (usually single spectroscopic confirmation following the astronomical telegram about \maria{the} transient discovery) we consider a supernova as \maria{a} SN~IIb candidate. 

\begin{figure}
    \begin{center}
        \includegraphics[width=\columnwidth]{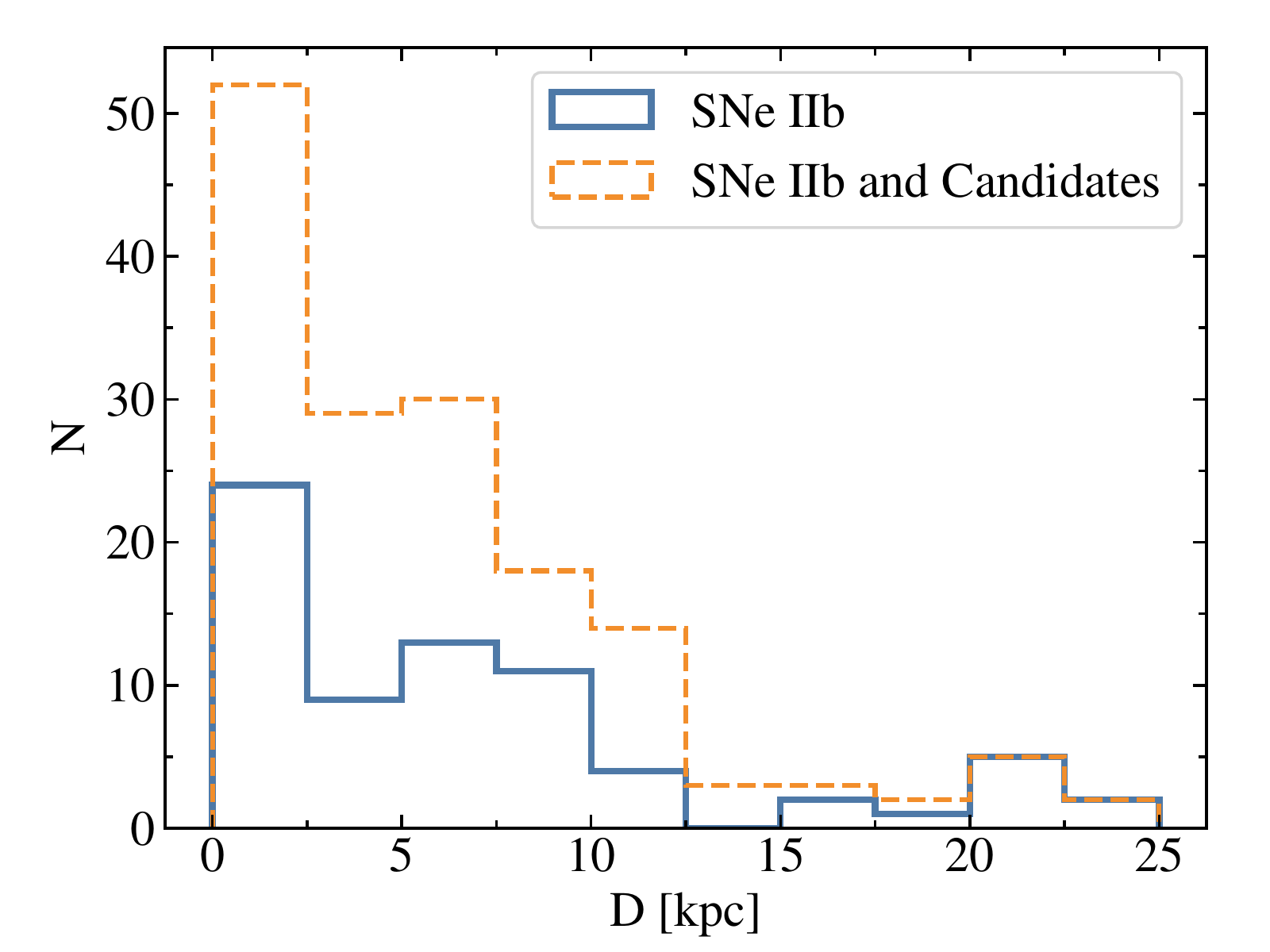}
        \caption{Histogram of \maria{the} supernovae distribution depending on \maria{the} projection of \maria{the} distance between SN and its host galaxy centre, $D$. The orange dashed line corresponds to all considered supernovae including confirmed SNe~IIb and candidates for SNe~IIb. The solid blue line corresponds to the distribution of confirmed SNe~IIb only.}
        \label{fig:hist}
    \end{center}
\end{figure}

First, we calculated the absolute galactocentric distance $D$ for each object as $D \simeq d_a\times \Theta$. The angle $\Theta$ is the angle between the supernova and \elena{the} host galaxy centre. The angular distance $d_a$ for flat $\Lambda$CDM cosmology with $\Omega_\Lambda = 0.7$ and $H_0 = 70$~km~s$^{-1}$~Mpc$^{-1}$ is
\begin{equation}
d_a = \frac{c}{H_0\times(1+z)} \int\limits_0^z{\frac{dz'}{\sqrt{(1-\Omega_\Lambda)\times(1+z')^3 + \Omega_{\Lambda}}}},
\end{equation}
where $z$ is \maria{the} redshift and $c$ is the speed of light. The distribution of Type IIb Supernovae by $D$ is presented in Fig.~\ref{fig:hist}. Most SNe~IIb, about 85~per~cent, are located inside a radius of 12~kpc. However, there is a local maximum near 20~kpc, which may be due to the fact that the radius of galaxies can vary widely. 

To perform a more accurate analysis we determined \maria{the} SN-host separation relative to the host size. To characterise the size of a galaxy we used a $D_{25}$ value, which is the major diameter measured to\elena{the $B$-passband} 25~mag~arcsec$^{-2}$ isophote. The $D_{25}$ values were extracted from the HyperLEDA extragalactic data base~\citep{Hyperleda}.

The full list of studied supernovae and the absolute and relative distances are summarised in Table~\ref{tab:full_list}: the first column is \maria{the} number in the list for easier searching, the second column consists of the supernova names starting with confirmed SNe~IIb, and continuing with SNe~IIb candidates. The equatorial coordinates (RA, Dec.) of supernovae and their host galaxies are presented in the third, fourth, fifth and sixth columns, respectively. The seventh column indicates \maria{the} redshift $z$. $D_{25}$ is given in column eight. The angle $\Theta$ expressed in arcsec is shown in the ninth column. Columns 10 and 11 contain the absolute distance $D$ in kpc and relative separation normalized to the size of the host galaxies, respectively.

\begin{figure}
    \begin{center}
        \includegraphics[width=\columnwidth]{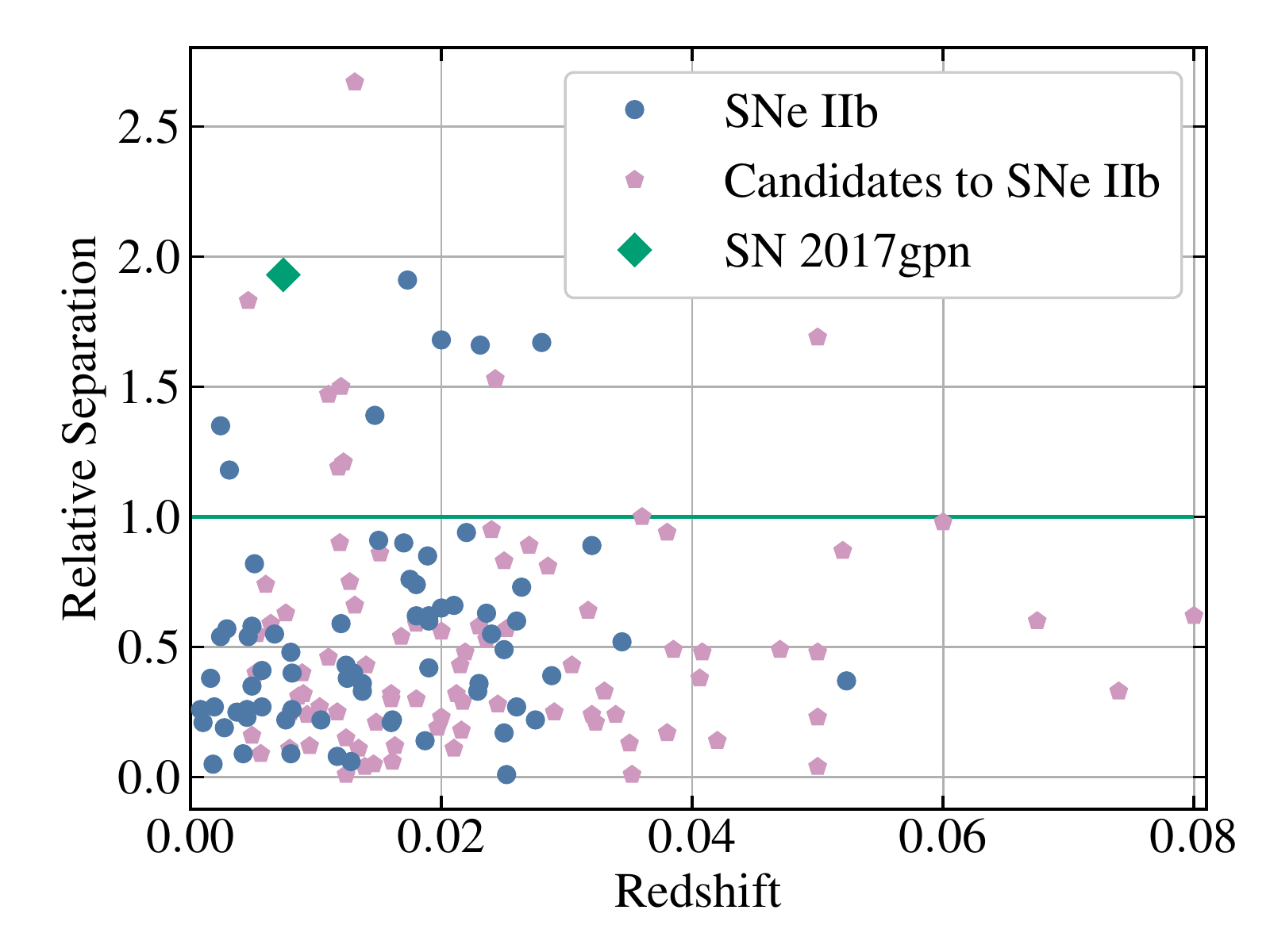}
        \caption{Relative separation between supernovae and their host galaxy centres as a function of redshift. Confirmed SNe~IIb are plotted in blue dots; candidates for SNe~IIb, in pink pentagons; and the studied SN~2017gpn is marked with \maria{the} green diamond. SNe above the green line are considered to be distant from the centres of their hosts.}
        \label{fig:rel_sep}
    \end{center}
\end{figure}

In Fig.~\ref{fig:rel_sep} we present \maria{the} relative separation between SNe and their host galaxies depending on the redshift. To evaluate how far away a supernova is, we chose a value of 1 for the relative separation, which is shown by the horizontal green line in Fig.~\ref{fig:rel_sep}. There are \elena{eight} SNe~IIb that lie above \maria{the} solid green line; we collect them into \maria{a} group of distant supernovae. SN~2017gpn is the most distant from the host galaxy centre \maria{among the} confirmed SNe~IIb. 

\begin{figure*}
    \begin{center}
        \includegraphics[width=2\columnwidth]{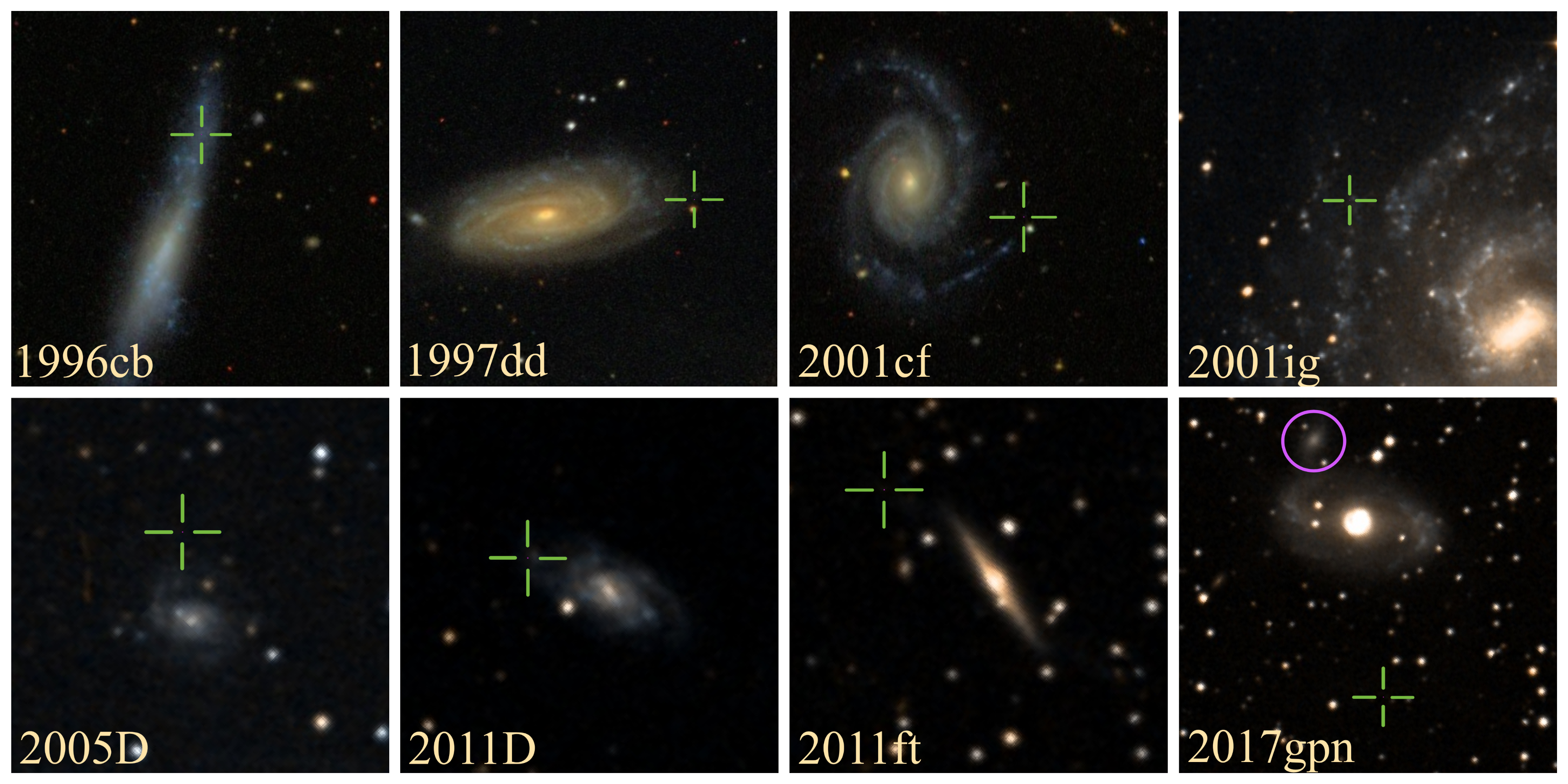}
        \caption{Optical images of supernovae distant from their host galaxy centres. SNe are marked by green crosses. All images were provided by SDSS~\citep{2017AJ....154...28B,1998AJ....116.3040G} and DSS. \thirdrev{The purple circle denotes the galaxy ZOAG~G134.74+13.65.}}
        \label{fig:collage}
    \end{center}
\end{figure*}

After that, we \maria{collected} images for all these distant SNe with the goal of investigating their unexpected \maria{location} (see Fig.~\ref{fig:collage}). \maria{The} majority of them are in continuations of spiral arms, e.g. supernovae 1997dd or 2001cf. \elena{Exceptions are supernovae 2011ft and 2017gpn, which are well outside the borders of their host galaxies.} We found Pan-STARRS1 images~\citep{2016Pan-starrs1,2016Pan-starrs2} for SN~2011ft in the $r$, $i$, $z$ and $y$ passbands where one can notice \maria{a} diffuse red object exactly at the SN~2011ft position, which can be associated with the host galaxy of SN~2011ft. 

In addition, we consider the object with the highest relative separation in Fig.~\ref{fig:rel_sep} (rel.~sep. is 2.67; see Table~\ref{tab:full_list}), SN~2017ati, \maria{a} candidate for type~IIb SNe. It turns out that this SN exploded in a system of interacting galaxies. Due to this interaction, \maria{a} region with \maria{a} high star-formation rate could be formed, and this explains the detection of \maria{the} core-collapse supernova far from the host galaxy disc. Therefore, SN~2017gpn is the only distant SNe that is not located in a region with a high star-formation rate.

According to the stellar evolution theory, the progenitor star of SN~IIb should be \maria{a} massive star with an initial mass of $\sim$30~$\Msun$. The fact that SN~2017gpn exploded far from a region with a high star-formation rate challenges this popular scenario. \maria{We have considered three different hypotheses to explain its location.}  

First, the progenitor of SN~2017gpn \maria{could be} a superspeed star. \cite{2005Brown} have discovered \maria{a} hypervelocity star SDSS~J090745.0+024507 with a mass of $\sim$4~$\Msun$ ejected from the Milky Way centre and left with a velocity of \elena{709~km~s$^{-1}$. If we presume that \maria{the} SN~2017gpn progenitor mass is about 30~$M_{\odot}$, the average lifetime of such a star will be $\sim$3~Myr calculated by the formula $ t_{life} \simeq \left(\frac{\Msun}{M_{star}}\right)^2 $. If it moves at a speed of 1000~km~s$^{-1}$~\citep{1988Hills},} it could move away from the centre of the host galaxy by $\sim$29~kpc during its lifetime. However, such a high velocity \maria{implies} that \maria{the} kinetic energy is $\sim3 \times 10^{50}$~erg\elena{; therefore an} effective mechanism of star acceleration is required. 

\begin{figure}
    \begin{center}
        \includegraphics[width=0.9\columnwidth]{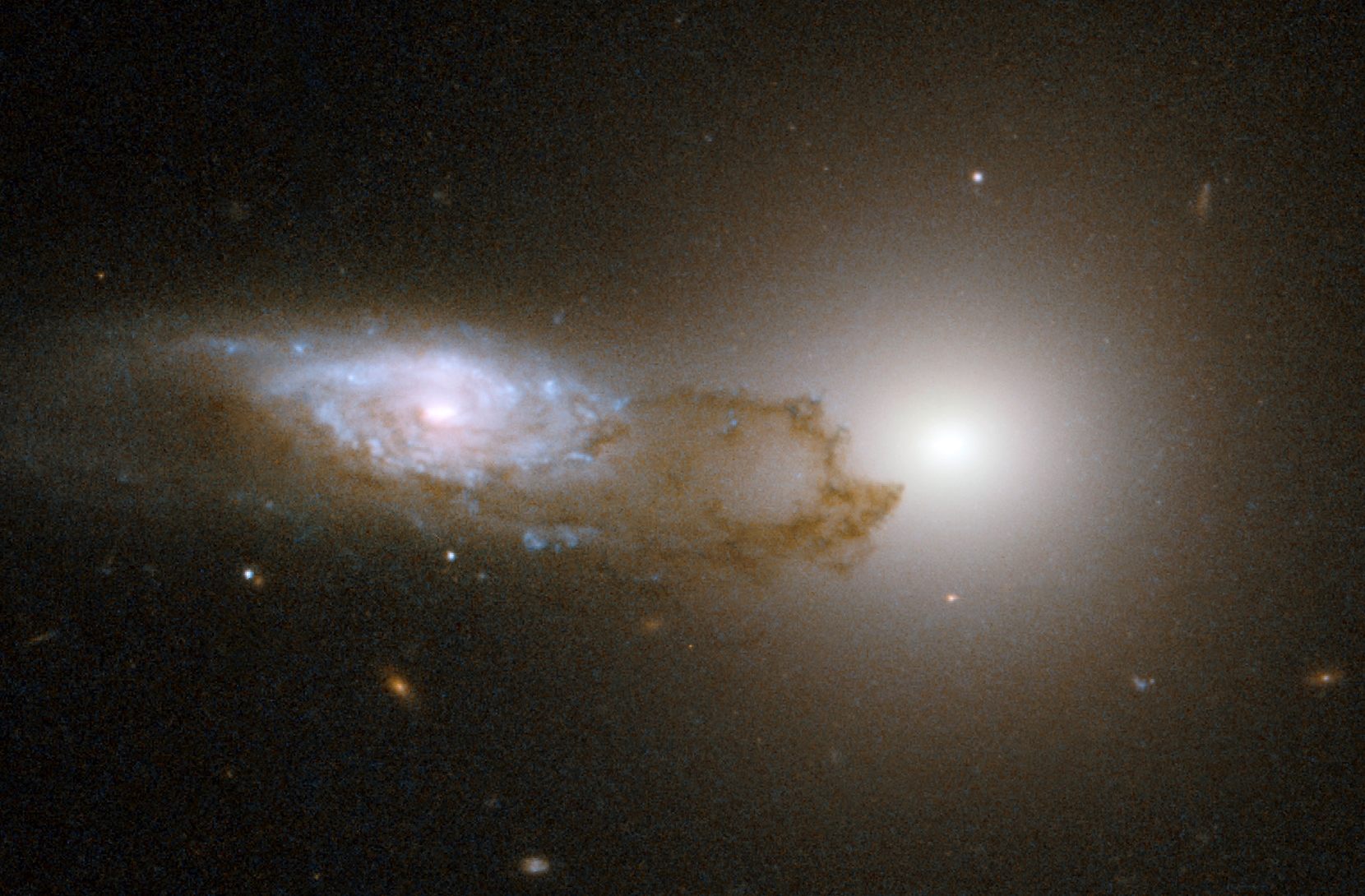}
        \caption{An image of AM~1316-241 obtained by \maria{the} Hubble Space Telescope~\citep{2001Keel}. The faint spiral arms are visible owing to the light from the background elliptical galaxy.}
        \label{fig:am1316}
    \end{center}
\end{figure}

The second hypothesis is that part of \maria{the} spiral arm of the host galaxy NGC~1343 is faint and therefore cannot be easily observed. For example, a similar situation is observed for the object AM~1316--241~(\citealt{2001Keel}; see Fig.~\ref{fig:am1316}). In this case we can see the faint spiral arm of \maria{the} galaxy only because it is illuminated by the light of a background elliptical galaxy. It is important that this part of \maria{the} spiral structure does not lie on the continuation of the bright spiral arm; therefore, a SN explosion there (in the absence of a ``lamp" behind) will appear to be outside the galaxy. 

The \maria{third} hypothesis is that the host galaxy of SN~2017gpn experienced \maria{an} interaction with other galaxies in the past. Tidal force destroyed the satellite galaxy and provided enough gas, which could condense far from \maria{the} NGC~1343 centre. Moreover, we can see the interaction between the galaxy ZOAG~G134.74+13.65 and \maria{the} SN~2017gpn host galaxy, which could also cause \maria{the} formation of gas clouds with a high star-formation rate \thirdrev{(see Fig. 12, panel 8)}.

\subsection{Connection with G299232}
Initially SN~2017gpn was considered as a possible optical counterpart of \maria{the} GW~event G299232 since it was discovered 2~d later in its error-box\footnote{\url{https://gcn.gsfc.nasa.gov/other/G299232.gcn3}}. If we assume that gravitational energy is released by a collapse, GW~events are expected from supernova explosions~\citep{1994Herant} and could be detected by \maria{the} LIGO/Virgo experiment~\citep{2019arXiv190803584T}. 

Nevertheless, the results of \maria{the} hydrodynamic modelling show that the explosion happened on Aug~20 ($\sim$3.5~d before the GW~alert) following the best-fitting model, or on Aug~17 for the additional model, i.e. $\sim$8~d before registration of G299232. \maria{G299232 is a low-significance event, so it could be a false signal; even if it is not, it is still implausible that SN~2017gpn could be associated with this alert}. Neither of our calculated models favor the electromagnetic counterpart of the gravitational event.

\section{Conclusions}
\label{sec:conclusion}
In this paper we have presented spectroscopic and photometric observations of \maria{the} Type~IIb Supernova 2017gpn and the results of the numerical modelling of its $B$, $R$ light curves with \maria{the} \textsc{STELLA} code. The best-fitting hydrodynamic model has the following parameter values: pre-SN radius $50~R_{\odot}$, pre-SN mass $3.5~\Msun$, mass of synthesized nickel totally mixed in the envelope $0.11~\Msun$, mass of \maria{the} compact remnant $1.41~\Msun$ (i.e. neutron star as a remnant), and energy of \maria{the} explosion $1.2 \times 10^{51}$~erg. \secrev{We also determined the ranges for these parameters by considering the dependence of the modelled light curves on each parameter while the others remain fixed. The obtained ranges are 3--4~$\Msun$ for the pre-SN mass, 20--70~$R_\odot$ for the pre-SN radius, 0.09--0.13~$\Msun$ for the mass of $\rm ^{56}Ni$, and, finally, (1.05--1.60)$ \times 10^{51}$~erg for $\rm E_{\rm exp}$.}

The study of Type~IIb Supernovae is an important part of \maria{the} exploration of the chemical composition of the Universe. The nucleosynthesis yields of CCSNe including SNe~IIb are characterized by strong contributions to the so-called alpha elements O, Ne, Mg, Si, S, Ar, Ca, and Ti~\citep{2018Thielemann} and \maria{the} heavy elements\maria{,} namely Ni, Co and Fe. 

According to the Open Supernova Catalog~\citep{SneSpace} only about a couple of dozen SNe~IIb have detailed photometry that allows the performance of reliable hydrodynamic modelling. \elena{Some of these SNe are considered in this paper and compared with SN~2017gpn taking into account a physically motivated classification of stripped-envelope SNe proposed by~\citet{PM2017}. In this classification SN~2017gpn belongs to the group IIb(I) which is characterized by strong hydrogen line profiles before maximum light, which weaken greatly over time, and $\rm H~\alpha$~P~Cygni profile dominated by the absorption component.} Analysis of \maria{the} hydrodynamic modelling results of different SNe~IIb shows that \maria{the} mass of synthesised $\rm ^{56}Ni$ is in the range 0.05--0.15~$\Msun$.

\maria{The} modelling results for SN~2017gpn are consistent with those for \elena{SNe~IIb considered, especially if we compare them with the modelling results for SN~2008ax which is of IIb(I) group according to~\citet{PM2017}.} These results together with the observational data presented here contribute to the study of \maria{the} Type IIb SN phenomenon, increasing the sample of well-studied SNe IIb. 

Finally, we considered three different hypotheses that could explain SN~2017gpn's distant \maria{location} relative to its host galaxy:
\begin{itemize}
    \item the progenitor of SN~2017gpn is a hypervelocity star ejected by NGC~1343 with an average speed more than 1000~km~s$^{-1}$;
    \item the progenitor exploded in a faint spiral arm of the host galaxy;
    \item the progenitor is formed in a region of interaction between \maria{the} host galaxy and another galaxy in the past. 
\end{itemize}
However, there is still a chance that the SN~2017gpn progenitor was not a massive star exploding for some reason far from regions of high star-formation rate. \maria{A} similar idea was proposed for \maria{the} Type~Ibn Supernova PS1-12sk by~\cite{2019Hosseinzadeh}. This question is open and challenges modern star-evolution models.

\section*{Acknowledgements}
EAB acknowledges support from a Russian Science Foundation grant 18-72-00159 for studying the question of the distant position of SN~2017gpn relative host galaxy centre. SIB and MVP acknowledge support from a Russian Science Foundation grant 18-12-00522 for supernova modelling with the \textsc{STELLA} code.
The authors acknowledge the support from the Program of Development of M.V.~Lomonosov Moscow State University (Leading Scientific School ``Physics of stars, relativistic objects and galaxies''). 
The authors are grateful to O.~I.~Spiridonova and the Zeiss-1000 staff for help with observations. We thank K.~L.~Malanchev \secrev{and M.~Sh.~Potashov} for helpful discussions. XW is supported by the National Natural Science Foundation of China (NSFC grants 11325313, 11633002, and 11761141001), and the National Program on Key Research and Development Project (grant no. 2016YFA0400803). We acknowledge the support of the staff of the Xinglong 2.16-m telescope. This work was partially supported by the Open Project Program of the Key Laboratory of Optical Astronomy, National Astronomical Observatories, Chinese Academy of Sciences. This research has made use of NASA's Astrophysics Data System Bibliographic Services and the following {\sc Python} software packages: {\sc NUMPY}~\citep{numpy}, {\sc MATPLOTLIB}~\citep{matplotlib}, {\sc ASTROPY}~\citep{astropy:2013, astropy:2018}.

\section{Data availability}
\thirdrev{The data underlying this article are available in the article.}


\bibliographystyle{mnras}
\bibliography{biblio} 

\appendix
\label{app}
\section{Table}
\onecolumn
    \begin{landscape}
    \begin{longtable}{lllllllrrrr}
        \caption{A complete list of confirmed Type IIb Supernovae and candidates to SNe~IIb.}
       \label{tab:full_list}
       \\
\hline
\textnumero & SN name & R.A.$_{\rm SN}$ & Dec.$_{\rm SN}$ & R.A.$_{\rm Host}$  & Dec.$_{\rm Host}$  & $z$  & $D_{25}$ [$''$] & $\Theta$ [$''$] & $D$ [kpc] & Rel. Sep. \\
\hline
1     & SN1987K                 & 12 43 41.17  & $+$16 23 44.9  & 12 43 42.63 & $+$16 23 36.2  & 0.0027 & 233.43  & 22.74  & 1.27  & 0.19     \\
2      & SN1993J                 & 09 55 24.77  & $+$69 01 13.7  & 09 55 33.17 & $+$69 03 55.1  & 0.0008 & 1312.66 & 167.57 & 2.78  & 0.26     \\
3      & SN1996cb                & 11 03 41.98  & $+$28 54 13.7  & 11 03 43.41 & $+$28 53 13.9  & 0.0024 & 92.93   & 62.68  & 3.11  & 1.35     \\
4      & SN1997dd                & 16 05 46.33  & $+$21 29 14.2  & 16 05 51.98 & $+$21 29 05.9  & 0.0147 & 114.33  & 79.30  & 23.77 & 1.39     \\
5      & SN1998fa                & 06 42 51.51  & $+$41 25 18.9  & 06 42 51.76 & $+$41 25 14.9  & 0.0250 & 57.3    & 4.89   & 2.46  & 0.17     \\
6      & SN2000H                 & 06 51 07.67  & $+$12 55 18.5  & 06 51 06.28 & $+$12 55 19.4  & 0.0130 & 101.89  & 20.34  & 5.40  & 0.40     \\
7      & SN2001cf                & 12 02 31.64  & $+$41 02 58.9  & 12 02 36.56 & $+$41 03 15.0  & 0.0200 & 68.89   & 57.94  & 23.48 & 1.68     \\
8      & SN2001gd                & 13 13 23.89  & $+$36 38 17.7  & 13 13 27.54 & $+$36 35 37.1  & 0.0029 & 586.34  & 166.50 & 9.99  & 0.57     \\
9      & SN2001ig                & 22 57 30.69  & $-$41 02 25.9  & 22 57 18.36 & $-$41 04 14.5  & 0.0031 & 300.71  & 176.76 & 11.33 & 1.18     \\
10     & SN2001Q                 & 11 25 19.77  & $+$63 43 15.6  & 11 25 19.05 & $+$63 43 45.4  & 0.0124 & 140.65  & 30.18  & 7.65  & 0.43     \\
11     & SN2002au                & 09 34 37.60  & $+$05 50 15.7  & 09 34 38.62 & $+$05 50 29.2  & 0.0180 & 65.79   & 20.34  & 7.44  & 0.62     \\
12     & SN2002eg                & 19 49 47.25  & $+$50 41 53.6  & 19 49 48.75 & $+$50 41 46.0  & 0.0260 & 53.48   & 16.15  & 8.45  & 0.60     \\
13     & SN2003bg                & 04 10 59.43  & $-$31 24 50.4  & 04 11 00.65 & $-$31 24 27.8  & 0.0046 & 101.89  & 27.47  & 2.61  & 0.54     \\
14     & SN2003cv                & 11 17 48.30  & $+$19 09 08.5  & 11 17 48.37 & $+$19 09 05.4  & 0.0288 & 16.53   & 3.25   & 1.88  & 0.39     \\
15     & SN2003ed                & 13 47 45.40  & $+$38 18 21.1  & 13 47 44.99 & $+$38 18 16.4  & 0.0045 & 52.26   & 6.74   & 0.63  & 0.26     \\
16     & SN2003gu                & 23 02 59.45  & $+$34 43 19.6  & 23 02 59.10 & $+$34 43 37.7  & 0.0190 & 60.00   & 18.61  & 7.17  & 0.62     \\
17     & SN2003ki                & 07 51 33.24  & $+$63 55 51.6  & 07 51 34.20 & $+$63 55 42.0  & 0.0250 & 46.57   & 11.50  & 5.79  & 0.49     \\
18     & SN2004be                & 10 00 19.47  & $-$24 48 13.8  & 10 00 19.30 & $-$24 48 08.0  & 0.0076 & 56.00   & 6.24   & 0.98  & 0.22     \\
19     & SN2004bi                & 10 47 37.45  & $+$26 18 12.0  & 10 47 39.37 & $+$26 17 41.5  & 0.0220 & 84.75   & 39.96  & 17.77 & 0.94     \\
20     & SN2004bm                & 10 52 35.33  & $+$22 56 05.5  & 10 52 35.75 & $+$22 56 02.8  & 0.0042 & 140.65  & 6.40   & 0.56  & 0.09     \\
21     & SN2004c                 & 11 27 29.76  & $+$56 52 48.4  & 11 27 31.89 & $+$56 52 36.2  & 0.0057 & 104.27  & 21.30  & 2.50  & 0.41     \\
22     & SN2004ex                & 00 38 10.19  & $+$02 43 17.2  & 00 38 12.38 & $+$02 43 42.6  & 0.0180 & 111.73  & 41.50  & 15.17 & 0.74     \\
23     & SN2004ff                & 04 58 46.19  & $-$21 34 12.0  & 04 58 47.12 & $-$21 34 09.9  & 0.0230 & 73.82   & 13.14  & 6.10  & 0.36     \\
24     & SN2004gj                & 11 30 59.63  & $+$20 28 06.8  & 11 31 00.66 & $+$20 28 08.6  & 0.0210 & 44.48   & 14.59  & 6.20  & 0.66     \\
25     & SN2005D                 & 07 26 57.36  & $+$20 22 53.4  & 07 26 57.12 & $+$20 22 15.5  & 0.0280 & 45.51   & 38.05  & 21.38 & 1.67     \\
26     & SN2005em                & 03 13 47.71  & $-$00 14 37.0  & 03 13 47.69 & $-$00 14 36.7  & 0.0252 & 95.09   & 0.42   & 0.22  & 0.01     \\
27     & SN2005H                 & 02 09 38.52  & $-$10 08 43.6  & 02 09 38.56 & $-$10 08 46.1  & 0.0128 & 80.94   & 2.57   & 0.67  & 0.06     \\
28     & SN2005U                 & 11 28 33.22  & $+$58 33 42.5  & 11 28 31.33 & $+$58 33 41.8  & 0.0010 & 143.93  & 14.80  & 0.31  & 0.21     \\
29     & SN2006ba                & 09 43 13.40  & $-$09 36 53.0  & 09 43 11.98 & $-$09 36 44.5  & 0.0190 & 106.70  & 22.66  & 8.73  & 0.42     \\
30     & SN2006bf                & 12 58 50.68  & $+$09 39 30.1  & 12 58 50.91 & $+$09 39 14.7  & 0.0240 & 57.30   & 15.77  & 7.63  & 0.55     \\
31     & SN2006el                & 22 47 38.50  & $+$39 52 27.6  & 22 47 37.39 & $+$39 52 44.8  & 0.0170 & 47.66   & 21.43  & 7.41  & 0.90     \\
32     & SN2006iv                & 11 48 12.35  & $+$54 59 14.6  & 11 48 11.32 & $+$54 59 30.2  & 0.0081 & 88.75   & 17.94  & 2.99  & 0.40     \\
33     & SN2006qp                & 14 42 30.65  & $+$28 43 25.9  & 14 42 33.24 & $+$28 43 35.2  & 0.0120 & 119.72  & 35.32  & 8.67  & 0.59     \\
34     & SN2006T                 & 09 54 30.21  & $-$25 42 29.3  & 09 54 28.64 & $-$25 42 11.8  & 0.0081 & 212.89  & 27.50  & 4.58  & 0.26     \\
35     & SN2007ay                & 08 17 14.85  & $+$01 12 06.9  & 08 17 15.73 & $+$01 12 23.0  & 0.0150 & 45.51   & 20.82  & 6.37  & 0.91     \\
36     & SN2008aq                & 12 50 30.42  & $-$10 52 01.4  & 12 50 29.39 & $-$10 51 15.7  & 0.0080 & 198.68  & 48.15  & 7.92  & 0.48     \\
37     & SN2008ax                & 12 30 40.80  & $+$41 38 14.5  & 12 30 36.41 & $+$41 38 37.4  & 0.0019 & 405.65  & 54.28  & 2.14  & 0.27     \\
38     & SN2008ay                & 12 55 26.36  & $+$52 16 15.5  & 12 55 24.90 & $+$52 16 03.5  & 0.0344 & 68.89   & 17.99  & 12.32 & 0.52     \\
39     & SN2008bo                & 18 19 54.34  & $+$74 34 20.9  & 18 19 46.42 & $+$74 34 06.2  & 0.0049 & 198.68  & 34.86  & 3.53  & 0.35     \\
40     & SN2008cx                & 00 56 45.90  & $-$09 54 19.0  & 00 56 42.66 & $-$09 54 50.1  & 0.0189 & 134.32  & 57.09  & 21.89 & 0.85     \\
41     & SN2008ie                & 02 43 20.80  & $+$04 58 19.1  & 02 43 22.27 & $+$04 58 06.2  & 0.0137 & 140.65  & 25.47  & 7.13  & 0.36     \\
42     & SN2009C                 & 23 13 42.84  & $+$49 40 47.2  & 23 13 43.95 & $+$49 40 35.7  & 0.0236 & 49.91   & 15.76  & 7.50  & 0.63     \\
43     & SN2009gk                & 21 44 27.28  & $+$14 53 57.3  & 21 44 28.76 & $+$14 53 59.2  & 0.0264 & 58.63   & 21.54  & 11.43 & 0.73     \\
44     & SN2009jv                & 09 40 57.83  & $+$47 37 04.0  & 09 40 58.19 & $+$47 37 13.3  & 0.0161 & 90.81   & 9.99   & 3.27  & 0.22     \\
45     & SN2009K                 & 04 36 36.77  & $-$00 08 35.6  & 04 36 37.35 & $-$00 08 37.0  & 0.0117 & 208.04  & 8.81   & 2.11  & 0.08     \\
46     & SN2009mk                & 00 06 21.37  & $-$41 28 59.8  & 00 06 19.92 & $-$41 29 59.6  & 0.0051 & 150.71  & 61.98  & 6.52  & 0.82     \\
47     & SN2010am                & 09 33 01.75  & $+$15 49 08.8  & 09 33 02.11 & $+$15 49 16.1  & 0.0200 & 27.20$^+$    & 8.96   & 3.63  & 0.65     \\
48     & SN2010cn                & 11 04 06.57  & $+$04 49 58.7  & 11 04 06.40 & $+$04 49 55.5  & 0.0260 & 30.30$^+$    & 4.09   & 2.14  & 0.27     \\
49     & SN2010ei                & 14 54 07.69  & $+$42 32 54.6  & 14 54 07.71 & $+$42 32 53.2  & 0.0187 & 20.20$^+$    & 1.42   & 0.54  & 0.14     \\
50     & SN2010ej                & 14 13 56.74  & $+$31 32 25.1  & 14 13 56.56 & $+$31 32 24.7  & 0.0523 & 12.54   & 2.34   & 2.38  & 0.37     \\
51     & SN2010ek                & 22 48 40.96  & $+$27 37 11.4  & 22 48 40.80 & $+$27 36 40.0  & 0.0320 & 70.49   & 31.47  & 20.12 & 0.89     \\
52     & SN2011bp                & 11 12 29.96  & $+$31 23 05.5  & 11 12 30.16 & $+$31 23 05.9  & 0.0275 & 23.89   & 2.59   & 1.43  & 0.22     \\
53     & SN2011D                 & 03 02 14.53  & $+$17 20 58.3  & 03 02 12.23 & $+$17 20 43.7  & 0.0231 & 43.47   & 36.02  & 16.80 & 1.66     \\
54     & SN2011dh                & 13 30 05.11  & $+$47 10 10.9  & 13 29 52.70 & $+$47 11 43.0  & 0.0016 & 828.23  & 156.49 & 5.19  & 0.38     \\
55     & SN2011ft                & 17 52 42.98  & $+$29 04 10.6  & 17 52 39.46 & $+$29 03 32.4  & 0.0173 & 62.83   & 59.91  & 21.07 & 1.91     \\
56     & SN2011fu                & 02 08 21.40  & $+$41 29 12.3  & 02 08 21.49 & $+$41 28 45.1  & 0.0190 & 90.81   & 27.22  & 10.49 & 0.60     \\
57     & SN2011hs                & 22 57 11.77  & $-$43 23 04.8  & 22 57 13.57 & $-$43 23 46.1  & 0.0057 & 337.40  & 45.72  & 5.37  & 0.27     \\
58     & SN2012P                 & 14 59 59.04  & $+$01 53 24.4  & 15 00 00.43 & $+$01 53 28.6  & 0.0045 & 181.20  & 21.26  & 1.98  & 0.23     \\
59     & SN2013ak                & 08 07 06.69  & $-$28 03 10.1  & 08 07 08.00 & $-$28 03 08.0  & 0.0037 & 140.65  & 17.47  & 1.34  & 0.25     \\
60     & SN2013bb                & 14 12 13.96  & $+$15 50 31.5  & 14 12 15.81 & $+$15 50 30.9  & 0.0175 & 70.49   & 26.70  & 9.50  & 0.76     \\
61     & SN2013df                & 12 26 29.33  & $+$31 13 38.3  & 12 26 27.09 & $+$31 13 24.8  & 0.0024 & 116.99  & 31.75  & 1.58  & 0.54     \\
62     & SN2014ds                & 08 11 16.45  & $+$25 10 47.4  & 08 11 15.92 & $+$25 10 45.7  & 0.0137 & 44.48   & 7.39   & 2.07  & 0.33     \\
63     & SN2015bi                & 14 32 15.31  & $+$26 19 32.0  & 14 32 15.19 & $+$26 19 36.2  & 0.0160 & 42.48   & 4.50   & 1.47  & 0.21     \\
64     & SN2016adj               & 13 25 24.12  & $-$43 00 57.9  & 13 25 27.60 & $-$43 01 08.8  & 0.0018 & 1542.24 & 39.69  & 1.48  & 0.05     \\
65     & SN2016gkg               & 01 34 14.46  & $-$29 26 25.0  & 01 34 18.24 & $-$29 25 06.6  & 0.0049 & 322.22  & 92.66  & 9.37  & 0.58     \\
66     & SN2017gpn               & 03 37 44.97  & $+$72 31 59.0  & 03 37 49.72 & $+$72 34 16.6  & 0.0074 & 143.93  & 139.25 & 21.20 & 1.93     \\
67     & ASASSN-14az             & 23 44 48.00  & $-$02 07 03.2  & 23 44 48.27 & $-$02 06 53.4  & 0.0067 & 38.74   & 10.6   & 1.46  & 0.55     \\
68     & ASASSN-14dq             & 21 57 59.97  & $+$24 16 08.1  & 21 57 59.82 & $+$24 15 59.7  & 0.0104 & 79.10   & 8.65   & 1.84  & 0.22     \\
69     & ASASSN-15bd             & 15 54 38.33  & $+$16 36 38.1  & 15 54 38.39 & $+$16 36 37.6  & 0.0080 & 22.81   & 1.00   & 0.16  & 0.09     \\
70     & PS15cjr                 & 02 38 07.29  & $+$01 23 29.2  & 02 38 07.57 & $+$01 23 18.1  & 0.0229 & 72.14   & 11.87  & 5.49  & 0.33     \\
71     & PTF11iqb                & 00 34 04.84  & $-$09 42 17.9  & 00 34 02.79 & $-$09 42 19.0  & 0.0125 & 157.82  & 30.33  & 7.75  & 0.38     \\
72     & SN2001ad$^*$                & 17 24 02.40  & $+$58 59 52.0  & 17 24 08.11 & $+$58 59 42.4  & 0.0110 & 61.40   & 45.15  & 10.17 & 1.47     \\
73     & SN2002hz$^*$                & 22 27 49.54  & $+$38 35 09.5  & 22 27 48.30 & $+$38 35 11.7  & 0.0180 & 99.58   & 14.70  & 5.38  & 0.30     \\
74     & SN2002jz$^*$                & 04 13 12.52  & $+$13 25 07.3  & 04 13 12.40 & $+$13 25 19.1  & 0.0052 & 60.00   & 11.93  & 1.28  & 0.40     \\
75     & SN2005by$^*$                & 13 45 46.91  & $+$22 05 46.8  & 13 45 45.62 & $+$22 05 18.4  & 0.0270 & 75.54   & 33.59  & 18.22 & 0.89     \\
76     & SN2005lr$^*$                & 07 11 39.03  & $-$26 42 20.2  & 07 11 40.45 & $-$26 42 17.9  & 0.0086 & 125.36  & 19.17  & 3.39  & 0.31     \\
77     & SN2006ss$^*$                & 14 20 27.46  & $+$35 11 42.7  & 14 20 26.50 & $+$35 11 19.1  & 0.0120 & 88.75   & 26.37  & 6.48  & 0.59     \\
78     & SN2008cw$^*$                & 16 32 38.27  & $+$41 27 33.2  & 16 32 38.00 & $+$41 27 33.0  & 0.0320 & 25.59   & 3.04   & 1.94  & 0.24     \\
79     & SN2008gx$^*$                & 10 15 32.95  & $+$74 12 59.1  & 10 15 32.22 & $+$74 13 13.1  & 0.0215 & 67.32   & 14.31  & 6.22  & 0.43     \\
80     & SN2009gj$^*$                & 00 30 28.56  & $-$33 12 56.0  & 00 30 21.89 & $-$33 14 43.3  & 0.0053 & 499.06  & 136.08 & 14.88 & 0.55     \\
81     & SN2009mg$^*$                & 06 21 44.86  & $-$59 44 26.0  & 06 21 38.91 & $-$59 44 24.0  & 0.0076 & 143.93  & 45.02  & 7.04  & 0.63     \\
82     & SN2009Z$^*$                 & 14 01 53.61  & $-$01 20 30.2  & 14 01 53.80 & $-$01 20 35.6  &        &         &        &       &          \\
83     & SN2010jr$^*$                & 05 19 35.80  & $-$32 39 28.2  & 05 19 35.81 & $-$32 39 27.9  & 0.0124 & 82.82   & 0.33   & 0.08  & 0.01     \\
84     & SN2011bv$^*$                & 13 02 53.57  & $-$04 02 36.0  &             &              &        &         &        &       &          \\
85     & SN2011cb$^*$                & 22 47 07.49  & $-$64 49 43.4  & 22 47 06.26 & $-$64 49 55.4  & 0.0079 & 261.91  & 14.34  & 2.33  & 0.11     \\
86     & SN2011ef$^*$                & 23 30 57.02  & $+$15 29 24.3  & 23 30 56.80 & $+$15 29 26.0  & 0.0134 & 64.29   & 3.61   & 0.99  & 0.11     \\
87     & SN2011ei$^*$                & 20 34 22.62  & $-$31 58 23.6  & 20 34 21.00 & $-$31 58 51.0  & 0.0093 & 280.64  & 34.29  & 6.55  & 0.24     \\
88     & SN2011hg$^*$                & 23 11 48.84  & $+$31 01 00.4  & 23 11 50.29 & $+$31 01 16.2  & 0.0236 & 92.93   & 24.44  & 11.64 & 0.53     \\
89     & SN2012cd$^*$                & 13 22 35.25  & $+$54 48 47.0  & 13 22 32.43 & $+$54 49 05.0  & 0.0118 & 51.07   & 30.30  & 7.32  & 1.19     \\
90     & SN2012dy$^*$                & 21 18 50.70  & $-$57 38 42.5  & 21 18 50.99 & $-$57 38 25.2  & 0.0103 & 128.28  & 17.46  & 3.69  & 0.27     \\
91     & SN2012fg$^*$                & 09 24 37.95  & $+$49 21 32.0  & 09 24 37.73 & $+$49 21 25.5  & 0.0163 & 114.33  & 6.85   & 2.27  & 0.12     \\
92     & SN2012hb$^*$                & 09 02 05.46  & $-$64 54 19.7  & 09 02 05.52 & $-$64 54 16.2  & 0.0056 & 80.94   & 3.52   & 0.41  & 0.09     \\
93     & SN2012hs$^*$                & 09 49 14.71  & $-$47 54 45.6  & 09 49 16.53 & $-$47 55 12.9  & 0.0064 & 111.73  & 32.86  & 4.33  & 0.59     \\
94     & SN2013bl$^*$                & 08 46 15.06  & $+$41 34 40.0  & 08 46 14.07 & $+$41 34 47.5  & 0.0304 & 62.83   & 13.40  & 8.15  & 0.43     \\
95     & SN2013cu$^*$                & 14 33 58.97  & $+$40 14 20.7  & 14 33 59.00 & $+$40 14 40.0  & 0.0252 & 67.32   & 19.30  & 9.80  & 0.57     \\
96     & SN2013ep$^*$                & 22 58 30.35  & $+$40 25 44.5  & 22 58 29.31 & $+$40 25 46.3  &        &         &        &       &          \\
97     & SN2013fq$^*$                & 19 59 07.95  & $-$55 55 46.6  & 19 59 06.40 & $-$55 55 41.6  &        &         &        &       &          \\
98     & SN2014cq$^*$                & 09 23 29.55  & $-$63 40 28.3  & 09 23 26.79 & $-$63 40 45.3  & 0.0110 & 109.18  & 25.02  & 5.64  & 0.46     \\
99     & SN2014ds$^*$                & 08 11 16.45  & $+$25 10 47.4  & 8 11 15.92  & $+$25 10 45.7  & 0.0137 & 44.48   & 7.39   & 2.07  & 0.33     \\
100    & SN2015au$^*$                & 22 30 59.42  & $-$13 59 56.1  & 22 30 59.91 & $-$14 00 12.8  & 0.0160 & 122.50  & 18.16  & 5.92  & 0.30     \\
101    & SN2015Y$^*$                 & 09 02 37.87  & $+$25 56 04.2  & 09 02 38.64 & $+$25 56 04.5  & 0.0080 & 84.75   & 10.39  & 1.71  & 0.25     \\
102    & SN2016avh$^*$               & 10 25 47.80  & $-$11 25 17.6  & 10 25 48.97 & $-$11 25 28.5  & 0.0380 & 43.47   & 20.36  & 15.35 & 0.94     \\
103    & SN2016bas$^*$               & 07 38 05.53  & $-$55 11 47.0  & 07 38 05.53 & $-$55 11 26.7  & 0.0090 & 128.28  & 20.30  & 3.75  & 0.32     \\
104    & SN2016bhr$^*$               & 14 25 20.58  & $+$32 28 55.9  & 14 25 20.59 & $+$32 28 56.5  & 0.0139 & 32.97   & 0.61   & 0.17  & 0.04     \\
105    & SN2016blq$^*$               & 11 08 55.51  & $-$29 01 26.4  & 11 08 55.52 & $-$29 01 25.5  &        &         &        &       &          \\
106    & SN2016blt$^*$               & 14 15 45.76  & $-$47 38 15.0  & 14 15 45.64 & $-$47 38 27.7  & 0.0160 & 79.10   & 12.76  & 4.16  & 0.32     \\
107    & SN2016bmd$^*$               & 07 20 24.3   & $+$32 51 01.2  & 07 20 24.60 & $+$32 50 58.8  &        &         &        &       &          \\
108    & SN2016dsb$^*$               & 01 58 59.71  & $-$32 22 18.5  & 01 59 00.57 & $-$32 22 25.2  &        &         &        &       &          \\
109    & SN2016exv$^*$               & 03 39 34.38  & $+$20 42 30.4  & 03 39 34.78 & $+$20 42 31.9  & 0.0212 & 36.15   & 5.81   & 2.49  & 0.32     \\
110    & SN2016hkn$^*$               & 02 08 34.23  & $+$29 14 11.1  & 02 08 34.37 & $+$29 14 02.6  & 0.0219 & 36.15   & 8.70   & 3.85  & 0.48     \\
111    & SN2016iyc$^*$               & 22 09 14.20  & $+$21 31 17.5  & 22 09 15.30 & $+$21 31 06.8  & 0.0127 & 49.91   & 18.71  & 4.86  & 0.75     \\
112    & SN2016iye$^*$               & 07 45 19.72  & $-$71 24 17.9  & 07 45 15.96 & $-$71 24 37.6  & 0.0180 & 90.81   & 26.67  & 9.75  & 0.59     \\
113    & SN2016M$^*$                 & 07 16 37.75  & $+$67 53 32.3  & 07 16 36.07 & $+$67 53 42.2  & 0.0360 & 27.43   & 13.71  & 9.81  & 1.00     \\
114    & SN2016U$^*$                 & 10 34 19.27  & $+$03 24 25.5  & 10 34 19.10 & $+$03 24 22.9  & 0.0740 & 21.78   & 3.64   & 5.12  & 0.33     \\
115    & SN2017ati$^*$               & 09 49 56.70  & $+$67 10 59.6  & 09 49 50.40 & $+$67 11 11.0  & 0.0131 & 28.72   & 38.38  & 10.27 & 2.67     \\
116    & SN2017cao$^*$               & 19 24 02.19  & $+$42 17 21.1  & 19 24 02.15 & $+$42 17 27.6  & 0.0200 & 23.34   & 6.52   & 2.64  & 0.56     \\
117    & SN2017dgd$^*$               & 16 45 38.967 & $+$01 37 19.7  & 16 45 39.02 & $+$01 37 13.1  &        &         &        &       &          \\
118    & SN2017eiy$^*$               & 23 49 28.27  & $-$30 25 04.7  & 23 49 28.64 & $-$30 25 14.8  & 0.0470 & 45.51   & 11.18  & 10.31 & 0.49     \\
119    & SN2017fek$^*$               & 20 21 47.44  & $-$10 43 53.3  & 20 21 47.70 & $-$10 43 46.0  & 0.0330 & 49.91   & 8.24   & 5.43  & 0.33     \\
120    & SN2017gfh$^*$               & 20 03 27.40  & $+$06 59 27.2  & 20 03 27.78 & $+$06 59 22.8  & 0.0245 & 51.07   & 7.17   & 3.54  & 0.28     \\
121    & SN2017gfz$^*$              & 00 12 51.89  & $-$32 43 53.0  & 00 12 51.80 & $-$32 44 02.0  & 0.0600 & 18.54   & 9.07   & 10.52 & 0.98     \\
122    & SN2017gkk$^*$               & 09 13 44.37  & $+$76 28 42.4  & 09 13 43.04 & $+$76 28 31.2  & 0.0049 & 154.22  & 12.13  & 1.23  & 0.16     \\
123    & SN2017gth$^*$               & 01 12 38.19  & $+$05 45 58.4  & 01 12 38.20 & $+$05 45 56.0  & 0.0380 & 28.72   & 2.40   & 1.81  & 0.17     \\
124    & SN2017hyh$^*$               & 07 10 41.07  & $+$06 27 41.4  & 07 10 40.48 & $+$06 27 13.0  & 0.0120 & 39.64   & 29.73  & 7.30  & 1.50     \\
125    & SN2017ixz$^*$               & 07 47 03.03  & $+$26 46 25.8  & 07 47 02.32 & $+$26 46 34.7  & 0.0240 & 27.43   & 13.02  & 6.30  & 0.95     \\
126    & SN2017iyd$^*$               & 11 46 25.00  & $+$01 59 33.1  & 11 46 24.70 & $+$01 59 39.6  & 0.0285 & 19.42   & 7.90   & 4.52  & 0.81     \\
127    & SN2017jbl$^*$               & 03 33 12.73  & $+$36 11 24.6  & 03 33 13.19 & $+$36 11 03.8  & 0.0151 & 49.91   & 21.53  & 6.63  & 0.86     \\
128    & SN2017jdn$^*$               & 10 23 45.51  & $+$53 06 20.5  & 10 23 46.90 & $+$53 06 28.0  & 0.0317 & 45.51   & 14.59  & 9.24  & 0.64     \\
129    & SN2017jo$^*$                & 09 57 36.150 & $-$22 10 23.91 & 09 57 36.41 & $-$22 10 30.7  &        &         &        &       &          \\
130    & SN2017mw$^*$                & 09 57 20.97  & $-$41 35 21.0  & 09 57 20.90 & $-$41 35 28.0  & 0.0117 & 57.30   & 7.04   & 1.69  & 0.25     \\
131    & SN2018arx$^*$               & 14 06 34.81  & $-$32 34 44.1  & 14 06 35.05 & $-$32 34 37.6  & 0.0339 & 58.63   & 7.17   & 4.85  & 0.24     \\
132    & SN2018bsg$^*$               & 10 10 28.16  & $+$02 13 48.8  & 10 10 27.86 & $+$02 13 41.6  & 0.0217 & 58.63   & 8.49   & 3.73  & 0.29     \\
133    & SN2018ddr$^*$               & 13 58 38.47  & $+$07 13 01.2  & 13 58 38.56 & $+$07 12 59.4  & 0.0146 & 82.82   & 2.24   & 0.67  & 0.05     \\
134    & SN2018dfg$^*$               & 14 06 34.70  & $-$05 27 02.9  & 14 06 34.89 & $-$05 27 10.7  & 0.0095 & 137.45  & 8.30   & 1.62  & 0.12     \\
135    & SN2018fcx$^*$               & 04 05 56.72  & $-$15 08 43.6  & 04 05 55.90 & $-$15 08 58.9  & 0.0250 & 46.57   & 19.37  & 9.75  & 0.83     \\
136    & SN2018fex$^*$               & 03 55 20.77  & $-$56 45 14.6  & 03 55 21.66 & $-$56 44 46.6  & 0.0243 & 37.86   & 28.94  & 14.18 & 1.53     \\
137    & SN2018fpb$^*$               & 23 59 42.80  & $+$34 20 39.9  & 23 59 42.96 & $+$34 20 42.6  & 0.0148 & 32.22   & 3.35   & 1.01  & 0.21     \\
138    & SN2018gj$^*$                & 16 32 02.31  & $+$78 12 40.9  & 16 32 39.20 & $+$78 11 53.5  & 0.0046 & 134.32  & 122.64 & 11.65 & 1.83     \\
139    & SN2018hhs$^*$               & 23 49 58.18  & $+$07 04 23.7  & 23 49 58.17 & $+$07 04 19.7  &        &         &        &       &          \\
140    & SN2018hqu$^*$               & 12 16 33.78  & $+$41 31 56.5  & 12 16 33.76 & $+$41 31 56.0  & 0.0500 & 28.72   & 0.55   & 0.54  & 0.04     \\
141    & SN2018hyw$^*$               & 08 20 17.38  & $+$20 52 32.2  & 08 20 16.57 & $+$20 52 30.3  & 0.0168 & 42.48   & 11.51  & 3.93  & 0.54     \\
142    & SN2018iuq$^*$               & 07 05 53.44  & $+$12 53 34.7  & 07 05 53.41 & $+$12 53 36.8  &        &         &        &       &          \\
143    & SN2018jak$^*$               & 09 59 18.19  & $+$34 53 43.8  & 09 59 18.13 & $+$34 53 53.3  & 0.0385 & 38.74   & 9.53   & 7.27  & 0.49     \\
144    & SN2018jee$^*$               & 07 23 14.632 & $+$56 31 30.45 & 07 23 14.45 & $+$56 31 29.6  &        &         &        &       &          \\
145    & SN2018mc$^*$                & 18 01 00.832 & $+$61 41 46.92 &             &              &        &         &        &       &          \\
146    & SN2018ow$^*$                & 02 51 04.410 & $+$09 06 44.32 & 02 51 04.80 & $+$09 06 64.3  &        &         &        &       &          \\
147    & SN2019abp$^*$              & 16 23 26.53  & $+$22 29 10.1  & 16 23 26.24 & $+$22 29 08.6  & 0.0376 &         &        &       &          \\
148    & SN2019abp$^*$               & 16 23 26.534 & $+$22 29 10.11 & 16 23 26.21 & $+$22 29 08.6  &        &         &        &       &          \\
149    & SN2019abp$^*$              & 16 23 26.534 & $+$22 29 10.11 & 16 23 26.21 & $+$22 29 08.6  &        &         &        &       &          \\
150    & SN2019ail$^*$               & 10 28 27.28  & $+$12 42 21.8  & 10 28 27.32 & $+$12 42 14.6  & 0.0323 & 68.89   & 7.22   & 4.66  & 0.21     \\
151    & SN2019aur$^*$               & 03 01 10.42  & $+$41 23 45.8  & 03 01 10.15 & $+$41 23 46.7  & 0.0124 & 43.47   & 3.17   & 0.80  & 0.15     \\
152    & SN2019bao$^*$               & 10 29 18.52  & $+$06 07 21.8  & 10 29 15.50 & $+$06 07 40.8  & 0.0119 & 109.18  & 48.88  & 11.90 & 0.90     \\
153    & SN2019bzo$^*$               & 15 55 34.45  & $+$26 54 54.8  & 15 55 34.46 & $+$26 54 54.0  & 0.0650 &         &        &       &          \\
154    & SN2019daf$^*$               & 13 47 48.12  & $+$72 03 00.4  & 13 47 48.56 & $+$72 02 59.8  & 0.0350 & 33.74   & 2.12   & 1.48  & 0.13     \\
155    & SN2019eev$^*$               & 09 57 05.86  & $+$08 04 10.17 & 09 57 05.86 & $+$08 04 10.17 &        &         &        &       &          \\
156    & SN2019eff$^*$               & 16 33 39.14  & $+$13 54 36.6  & 16 33 39.23 & $+$13 54 23.6  & 0.0500 & 15.42   & 13.07  & 12.77 & 1.69     \\
157    & SN2019fco$^*$               & 10 32 04.72  & $+$46 55 03.6  & 10 32 04.34 & $+$46 55 03.8  & 0.0406 & 20.33   & 3.90   & 3.13  & 0.38     \\
158    & SN2019fks$^*$               & 21 41 16.50  & $-$16 53 23.0  & 21 41 16.80 & $-$16 53 23.0  & 0.0500 & 18.12   & 4.31   & 4.21  & 0.48     \\
159    & SN2019gaf$^*$               & 20 36 55.23  & $+$02 48 24.6  & 20 36 54.90 & $+$02 48 14.0  & 0.0060 & 31.49   & 11.70  & 1.45  & 0.74     \\
160    & SN2019iij$^*$               & 14 56 33.53  & $-$25 50 10.9  & 14 56 33.62 & $-$25 50 06.9  & 0.0290 & 33.74   & 4.18   & 2.43  & 0.25     \\
161    & SN2019ltw$^*$               & 16 18 38.54  & $+$21 58 22.9  & 16 18 38.53 & $+$21 58 24.0  & 0.0161 & 35.33   & 1.11   & 0.36  & 0.06     \\
162    & SN2019rn$^*$                & 02 17 59.61  & $+$14 32 00.4  & 02 17 59.65 & $+$14 32 38.2  & 0.0131 & 114.33  & 37.80  & 10.12 & 0.66     \\
163    & SN2019xt$^*$                & 14 11 55.741 & $-$00 50 11.77 & 14 11 55.78 & $-$00 50 12.0  &        &         &        &       &          \\
164    & SNhunt268$^*$               & 01 14 26.82  & $+$42 33 18.4  & 01 14 26.27 & $+$42 33 22.6  & 0.0197 & 77.29   & 7.39   & 2.95  & 0.19     \\
165    & ASASSN-15qz$^*$             & 01 25 36.09  & $-$41 27 55.8  & 01 25 35.80 & $-$41 27 55.5  & 0.0216 & 36.15   & 3.27   & 1.43  & 0.18     \\
166    & CSS151130$^{*1}$            & 01 42 58.44  & $+$27 34 10.5  & 01 42 57.60 & $+$27 34 53.0  &        &         &        &       &          \\
167    & DES16S1kt$^*$               & 02 51 07.54  & $+$00 01 33.0  & 02 51 07.77 & $+$00 01 30.8  & 0.0675 & 13.75   & 4.09   & 5.29  & 0.60     \\
168    & iPTF13efs$^*$               & 07 55 26.19  & $+$52 48 17.9  & 07 55 25.98 & $+$52 48 24.7  & 0.0408 & 29.39   & 7.06   & 5.69  & 0.48     \\
169    & LSQ12hbo$^*$                & 10 56 16.00  & $-$20 51 32.0  & 10 56 12.97 & $-$20 51 10.1  & 0.0122 & 79.10   & 47.79  & 11.93 & 1.21     \\
170    & LSQ12htu$^*$                & 10 11 37.11  & $-$07 23 11.6  & 10 11 37.68 & $-$07 23 15.2  & 0.0520 & 21.29   & 9.21   & 9.34  & 0.87     \\
171    & LSQ13bca$^*$                & 21 16 16.39  & $-$20 30 48.9  & 21 16 16.80 & $-$20 30 54.0  & 0.0800 & 25.01   & 7.69   & 11.62 & 0.62     \\
172    & LSQ14hj$^*$                 & 13 25 08.05  & $-$32 37 32.8  & 13 25 07.90 & $-$32 37 31.0  & 0.0500 & 22.81   & 2.61   & 2.55  & 0.23     \\
173    & LSQ15rw$^*$                 & 14 32 31.33  & $-$13 39 27.4  & 14 32 31.19 & $-$13 39 26.0  & 0.0210 & 43.47   & 2.47   & 1.05  & 0.11     \\
174    & OGLE16ekf$^*$               & 04 37 36.66  & $-$71 48 17.4  &             &              &        &         &        &       &          \\
175    & PS1-14od$^*$                & 03 21 06.23  & $-$07 16 57.4  & 03 21 06.08 & $-$07 16 56.8  & 0.0200 & 19.87   & 2.31   & 0.94  & 0.23     \\
176    & PS15apj$^*$                 & 18 28 58.24  & $+$22 54 10.6  & 18 28 57.36 & $+$22 54 11.0  & 0.0140 & 56.00   & 12.17  & 3.48  & 0.43     \\
177    & PS15bgt$^*$                 & 22 46 05.04  & $-$10 59 48.4  & 22 46 03.70 & $-$11 00 04.3  & 0.0089 & 125.36  & 25.34  & 4.63  & 0.40     \\
178    & PS15bqc$^*$                 & 17 04 32.29  & $+$01 20 58.5  & 17 04 32.26 & $+$01 20 47.7  & 0.0230 & 37.00   & 10.81  & 5.02  & 0.58     \\
179    & PTF10htz$^*$                & 13 08 37.52  & $+$79 47 13.2  & 13 08 37.55 & $+$79 47 13.3  & 0.0352 & 28.06   & 0.13   & 0.09  & 0.01    \\
\hline
\end{longtable}
$^+$ The value of D25 is measured as isophotal level of 25 mag~arcsec$^{-2}$ in the SDSS g-band~\citep{2012Hakobyan, 2016Hakobyan}.

$^*$ Candidates to Type IIb Supernovae.

$^1$ CSS151130:014258+273410
\end{landscape}

\bsp	
\label{lastpage}
\end{document}